\documentclass[]{interact}
\usepackage{subfigure}% Support for small, `sub' figures and tables
\usepackage{amsmath}
\usepackage{graphicx}
\usepackage{dcolumn}
\usepackage{soul,color}
\usepackage{bm}
\usepackage{times}
\usepackage[numbers,sort&compress,merge]{natbib}% Citation support using natbib.sty
\renewcommand\bibfont{\fontsize{10}{12}\selectfont}% Bibliography support using natbib.sty
\renewcommand\bibnumfmt[1]{(#1)}% Parentheses enclose ref. numbers in list using natbib.sty

\theoremstyle{plain}% Theorem-like structures

\theoremstyle{definition}

\theoremstyle{remark}

\begin{document}

\title{Multiparameter estimation, lower bound on quantum Fisher information and  non-Markovianity witnesses of noisy two-qubit systems }

\author{
\name{H. Rangani Jahromi$^{1}$\thanks{ Email: h.ranganijahromi@jahromu.ac.ir},  M. Amini$^{2}$, M. Ghanaatian$^{2}$}
\affil{$^{1}$Physics Department, Faculty of Sciences, Jahrom
University, P.B. 7413188941,  Jahrom, Iran.
}
\affil{
$^{2}$ Department of Physics, Payame Noor University (PNU), P.O. Box 19395-3697, Tehran, Iran.
}
}

\maketitle

\begin{abstract}
By using the quantum Fisher
information (QFI), we address the process of \textit{single}-parameter estimation in the presence of bosonic as well as fermionic environments and protection of information against the noise. In particular, the quantum interferometric power (IP) of the evolved state of the system is uncovered as an important lower bound for the QFIs of initially encoded parameters. Moreover, we unveil new witnesses of non-Markovianity, 
that can be used to detect efficiently  the memory effects and backflow of information from the
environment to the system. 
On the other hand,  we also investigate the \textit{multiparameter} estimation of initial parameters encoded into the quantum state of a two qubit system and obtain analytical formula of the corresponding QFI matrix.  In particular, the corresponding quantum
Cramer-Rao bounds in both single and multiparameter estimations are analysed. 
In addition,  we illustrate that the 
quantum \textit{coherence} and \textit{purity} of the evolved state of the probes are two key elements in realizing optimum multiparameter estimation.

\end{abstract}

 \maketitle
\maketitle
\small
\section{Introduction\label{introduction}}

Direct measurements of many quantities in physics are not
possible, either in principle or due to experimental limitations. In particular, this is  true for quantum systems
where their variables such as purity and entanglement 
are associated with no quantum observables.
 Under these conditions, the values of the parameter are usually inferred from a set of
 observables or a set
of indirect measurements of a different observable.
This procedure is addressed in quantum estimation theory by using a measure called the quantum Fisher information (QFI) \cite{M.G.A. Paris,Holevo2011,Zhang Jiang,Wei Zhong,J. Ma,Yao Yao,V. GiovannettiPRL,RanganiAOP,Jian Ma,RanganiOPTC,Toth2013,RanganiAOP2,RanganiJMO,J. Liu11,RanganiQIP4,RanganNEW2}. In  the formalism  of
quantum sensing and metrology, the objective is to find
the fundamental precision bounds of parameter estimations and
the optimal measurement strategies saturating them.
 Early
work in this field outstandingly focused on single parameter estimations \cite{Giovannetti2011}, whose quantum enhanced limit has been
proven to always be attainable \cite{M.G.A. Paris}.
 However, applications of quantum 
metrology to microscopy, optical, electromagnetic, and gravitational field imaging usually demand multiple parameter estimations \cite{Sidhu2018}. 
Hence, recent years have seen a surge of interest in exploring 
 enhancement of quantum metrology in the case of simultaneous estimation of multiple parameters \cite{Monras2011, Genoni2013,Humphreys2013,Crowley2014,Vidrighin2014,Y. Yao2014,Baumgratz2016,Knott2016}.
 Multi-parameter quantum
enhanced sensing has provided a novel strategy for studying the information processing capabilities of multipartite
or multimode quantum correlated states and measurements.

Recently, inspired by the seminal work of Baumgratz
et al. \cite{Baumgratz2014}, there is increasing interest in 
quantitative and axiomatic studies of coherence \cite{Lostaglio2015,Yuan2015,Shao2015,Winter2016,Chitambar2016,Napoli2016,Marvian2016,Rana2016}.
Quantum coherence is one of the old but always significant concepts in quantum mechanics, and
 now it has been investigated as a necessary resource for quantum information processing and quantum
 metrology. Because QFI is  mathematically related to some
 other functions previously proposed to quantify the quantum coherence, such as the relative entropy \cite{Frieden2004}, fidelity based on the distance measurement \cite{Braunstein1994}, 
 and the skew information \cite{Petz2014}, etc., it is logically reasonable \cite{Luis2012,Feng2017} to use the
 QFI for quantifying the quantum coherence. More interestingly,  the proposal according to the QFI to quantify quantum coherence is experimentally testable, different from most of the pure axiomatic functions proposed
  previously, as the
 lower and upper bounds of the QFI are practically measurable. Because QFI plays a fundamental role in parameter
 estimation, it is desirable to investigate whether the coherence
 measure provides any insight into multiparameter quantum metrology.

 \par
One of the fundamental concepts playing an important role in quantum metrology is non-Markovianity \cite{Breuer2002}  often characterized in terms of the backflow of
information from environment to system or the presence of the environment memory. A number of non-Markovian measures and witnesses have
been recently proposed \cite{Rivas2014}. 
Among the most significant ones, let us mention
those based on the deviation of the dynamical maps from divisible CPTP maps \cite{Rivas2010,Wolf2008} and those based on
channel capacities \cite{Bylicka2014}, the nonmonotonicity
of the trace distance or distinguishability \cite{Breuer2009}, entanglement \cite{Rivas2010}, and quantum mutual information \cite{Luo862012}.
 Other significant works to quantify the non-Markovianity
include the flow of QFI \cite{X2010}, volume
of Gaussian states \cite{Lorenzo2013}, local quantum uncertainty \cite{Z. He2014}, and coherence \cite{ Chanda2016,Zhi He2017}. Moreover, in Ref. \cite{Rajagopal2010}, it has been shown that  fidelity between dynamical time-evolved states  is a witness of non-Markovianity such that negative \textit{fidelity
difference} reveals the non-Markovian feature. Also, the fidelity associated with the initial state
and the dynamically evolved state was shown to be larger in the non-Markovian evolution compared
to that in the corresponding Markovian case. Hence, it is interesting to investigate more the role of fidelity as a witness of non-Markovianity. 
On the other hand, in Ref.  \cite{Dhar2015}, after attaching 
an ancilla A to the (principal) system S, the authors introduced the quantum interferometric power (IP) \cite{Girolami2014,Lecture}, defined in terms of the minimal QFI obtained by local unitary evolution of the ancilla in an interferometry process, to characterize the non-Markovianity interaction of
the principal system with its environment. Now a question arises: Can IP reveal the non-Markovian behaviour of the system without using an ancilla?

 \par 
 
 A larger QFI means that we can estimate
 the parameter with a higher precision.
 On the other hand, finding  a lower
 bound to the QFI, we can check whether or not one can obtain a minimum precision in the process of estimation. 
 S. Luo \cite{Luo2003PAMS,LuoPRL2003} showed   that in the unitary evolution the
 \textit{Wigner-Yanase skew information}  is majorized by the
 QFI associated with the phase parameter. Besides, in Ref. \cite{ChaoZhang2017} it has been experimentally extracted a lower bound to metrologically useful
    asymmetry and entanglement of a two-qubit system in an optical setup, by measuring its speed during a unitary evolution.
 Moreover,  it has been proven that \cite{Girolami2013,Lecture,Girolami2014}
 quantum correlations, measured by the local quantum uncertainty (LQU) or IP, that \textit{ were already present} in the initial (input) state, 
 guarantee a minimum sensitivity in the protocol of optimal
 phase estimation of local unitary evolution. We remark that the QFIs are obtained from
 the output estimation data, while the LQU or IP is measured on the
 input probe states.
On the other hand,
for few special
open quantum systems, i.e.,  two coupled qubits interacting with the
independent non-Markovian Lorentzian form environments \cite{Shao2018} and two modes of a Dirac field described by relatively accelerated parties \cite{RanganNEW}, it has been illustrated that the  LQU of the evolved state  is majorized by  the QFI associated with the phase parameter encoded into the initial state of the system.
 It is also interesting to
investigate how  the evolved state IP  of the open quantum system  can be employed to investigate the multi-parameter estimation and
metrological scaling associated with the initial state of the system.

\par In this paper we discuss about independent and simultaneous estimations of the entanglement as well as mixedness parameters of initial state of the system interacting with bosonic and fermionic (spin) environments. Removal of   the decoherence effects and  realization of the \textit{QFI trapping}, playing important roles  in quantum communication protocols, are also addressed.  We show in particular that the amount of evolved state quantum correlations as quantified by IP, guarantee a minimum precision, as quantified by the QFI, in the  estimation protocol. 
Moreover,  we find that the bosonic environment is robust against backflow of information to the system while the non-Markovian dynamics can  occur in the presence of spin environment. Hence, we explore the evolved state IP, LQU and fidelity as three novel witnesses of non-Markovianity such that the \textit{collapse-revival} behaviour of the fidelity and positive time-derivatives of the IP and LQU can be used to detect the intervals that the non-Markovianity occurs.  On the other hand, by analytical calculation  of the QFIM associated with
the evolution of the input  state, the  uncertainties have been derived and analysed through
the QCRBs for joint  estimation of the parameters.
 In addition, 
the role of coherence and purity in the process of multiparameter estimation is addressed.

  \par This paper is organized as follows. In Sec. \ref{pre}, a brief review of the multi-parameter estimation theory and the IP are presented. The models and the reduced density matrices   are introduced in Sec. \ref{Model}. In Sec. \ref{Single}, the problem of single parameter estimation and introduction of the IP as its lower bound  are analysed completely.  In Sec. \ref{flow},
   our  new witnesses of non-Markovianity are presented. Moreover, the multi-parameter estimation problem is discussed completely in Sec. \ref{Multi}. 
   Finally, Sec. \ref{conclusion}
    is devoted to the discussion and conclusion.

\section{The Preliminaries}\label{pre}

\subsection{Quantum parameter estimation}

Let $ \Phi_{\lambda} $  be a quantum channel depending on a set of parameters $ \lambda = (\lambda_{1}, . . . ,\lambda_{n}) $ that we intend to estimate them by sending an
input quantum probe $ \rho $ and then measuring  output $ \rho_{\lambda}=\Phi_{\lambda}(\rho) $.
The measurements correspond
to a positive operator valued measure (POVM), that is, a set
of positive operators  $\{\varPi_{x}\} $ such that $ \sum\limits_{x}^{}\varPi_{x}\varPi_{x}^{\dagger}=\mathcal{I} $ and
such that  $ p(x|\lambda)\equiv \prod\limits_{i}^{} p(x_{i}|\lambda)= \text{Tr}[\varPi_{x}\rho_{\lambda}]  $ denotes the probability distribution
for  results $ x=(x_{1},...,x_{M}) $ of the measurements performed M-times.
. Moreover, according to obtained results of the measurement, parameters are estimated
by using an estimator   $\tilde{\lambda}(x)=(\tilde{\lambda} _{1}(x),,...,\tilde{\lambda} _{n}(x)) $.
We say that  $\tilde{\lambda}(x) $
is an unbiased estimator if $ E(\tilde{\lambda})\equiv\sum\limits_{x}^{}p(x|\lambda)\tilde{\lambda}(x)=\lambda $ [i.e., its expected value
coincides with the true value of the parameter(s)] \cite{Brida2011,Ciampini2015}. 
The Cram\'{e}r-Rao
bound declares that, for all unbiased estimators $ \tilde{\lambda} $, the 
covariance matrix with elements defined as $\text{Cov}[\tilde{\lambda}]_{ij}=E[\tilde{\lambda}_{i}\tilde{\lambda}_{j}]-E[\tilde{\lambda}_{i}]E[\tilde{\lambda}_{j}] $
satisfies \cite{Humphreys2013,Vidrighin2014,Y. Yao2014}

          \begin{equation}\label{CRAMER}
\text{Cov}[\tilde{\lambda}]\geq \dfrac{1}{M}\textbf{I}(\lambda)^{-1},
          \end{equation}
where M represents the number of experimental runs (M = 1 for definiteness); and where $ \textbf{I}^{-1} $ denotes the inverse of the Fisher information (FI) matrix whose elements are given by \cite{Ciampini2015}

          \begin{equation}\label{CFIM}
\text{I}_{ij}=\sum\limits_{x}^{}\dfrac{1}{p(x|\lambda)}\dfrac{\partial p(x|\lambda)}{\partial \lambda_{i}} \dfrac{\partial p(x|\lambda)}{\partial \lambda_{j}}.
          \end{equation}
The classical FI is further bounded
by the quantum Fisher information  matrix (QFIM) $ \textbf{F} $ via the
matrix inequality $ \textbf{F}\geq \textbf{I}  $ \cite{Ivanov2018}. In order to compute the QFIM, we should first obtain the Symmetric Logarithmic Derivative (SLD) $ L_{\lambda_{j}} ~~ ( j =
1,2,..., p)$ satisfying the operator equation

         \begin{equation}\label{SLD1}
        \frac{\partial \rho(\lambda)}{\partial \lambda_{j}}=\dfrac{1}{2}[L_{\lambda_{j}}\rho(\lambda)+\rho(\lambda)L_{\lambda_{j}}].
         \end{equation}  
where $ \rho $ is the density operator. Then, the  corresponding QFIM
elements are given by

       \begin{equation}\label{QFIM}
      F_{ij}=\frac{1}{2}\text{Tr}[\rho(L_{\lambda_{i}}L_{\lambda_{j}}+L_{\lambda_{j}}L_{\lambda_{i}})].
         \end{equation}
The QFI gives the ultimate precision
in the multi-parameter estimation quantified by the quantum
Cramer-Rao bound (QCRB) 

          \begin{equation}\label{CRAMER}
\text{Cov}[\tilde{\lambda}]\geq \textbf{F}(\lambda)^{-1},
          \end{equation}
In the special case where each parameter is estimated \textit{independently} (i.e., single-parameter estimation), the inequality reads \cite{Brida2011,Braunstein1994}

         \begin{equation}\label{CRAMERsingle}
(\delta \lambda_{j})_{i}\geq \dfrac{1}{F_{jj}},
          \end{equation}
where $ \delta \lambda_{j}\equiv \text{Var}(\tilde{\lambda}_{j})\equiv E(\tilde{\lambda}_{j}^{2})-E(\tilde{\lambda}_{j})^{2} $ and 
 $ F_{jj}\equiv F(\lambda_{j}) $ represents the
QFI  of parameter $ \lambda_{j} $ given by

          \begin{equation}\label{QFImohem}
F(\lambda_{j}) =\text{Tr}[\rho(\lambda) L_{\lambda_{j}}^{2}].
          \end{equation}

        In the general case, i.e., \textit{simultaneous} estimation of parameters, the inequality for the variance of each  parameter is obtained as
        
                \begin{equation}\label{CRAMERmulti}
        (\delta \lambda_{j})_{s}\geq [\textbf{F}(\lambda)^{-1}]_{jj},
                  \end{equation}
        It is  clear to see that if there exists nonzero off-diagonal elements in the QFI matrix, the uncertainty bounds for
        simultaneous  estimation of parameters may be
        different from the independent cases.
        \par
        Taking
        the trace of both sides of Eq. (\ref{CRAMER}), we obtain  a lower bound on the total
        variance of all the parameters that should be estimated \cite{Humphreys2013}
        
                  \begin{equation}\label{CRAMERtotal}
                \delta \equiv \sum\limits_{j}^{} (\delta \lambda_{j})_{s}\equiv \text{Tr}\big(\text{Cov}[\tilde{\lambda}]\big) \geq \text{Tr}\big(\textbf{F}(\lambda)^{-1}\big),
                          \end{equation} 
         For a single parameter, using the eigenvectors of the SLD operator as the POVM \cite{Ragy2017}, we find that the
        equality  $\text{I}=\text{F}$ can always be achieved and hence the QCRB (\ref{CRAMERsingle}) may be saturated. 
       However, simultaneous estimation of parameters  in a single metrology protocol may be
        more challenging procedure than the individual estimation of them.
 Because the optimal measurement for a given
parameter is formed from projectors corresponding to the
eigenbasis of the SLD, we can immediately conclude that if
$ \forall ~ (\lambda_{j},\lambda_{k} ) \in \lambda:~[L_{\lambda_{j}},L_{\lambda_{k}}]=0  $ then there is a single eigenbasis for all SLDs and thus a common measurement optimal from the point of
view of extracting information on all parameters simultaneously. Of course, it is only a sufficient but not a necessary condition. 
  In fact, if the SLDs do not commute,
  this does not necessarily imply that it is impossible to simultaneously extract information on all parameters
  with precision matching that of the separate scenario for
  each \cite{Ragy2017}.

\subsection{Quantum interferometric power}\label{IPsec}

A significant phase estimation scenario is implemented by estimation through interferometric measurements \cite{Lecture}. 
A bipartite
system AB prepared into the input state $ \rho_{AB} $ is injected through a two-arm channel. Subsystem A experiences  phase shift $ U_{A}=e^{-iH_{A}^{\Lambda}\theta} $, generated by  Hamiltonian $ H_{A}^{\Lambda} $ having
non-degenerate spectrum $\Lambda  $, while subsystem B is completely unaffected.
 This restriction is applied in order to understand the role of
non-classical correlations in this scenario. 
The phase $ \theta $, being not directly measurable,  denotes
the unknown parameter we intend to estimate.
 Moreover, 
it is supposed that the estimation is blind, i.e.,  only the spectrum of the Hamiltonian, generating the encoded phase,  is
known during the input preparation of the phase estimation. There is no more information about  eigenbasis of the Hamiltonian .
It can be shown that, in this scenario, a quantifier for the worst-case precision is a \textit{bona fide}
measure of non-classical correlations.  The worst-case QFI for a given state is given by \cite{Girolami2014}

          \begin{equation}\label{IP}
\text{IP}_{A}^{\Lambda}(\rho_{AB})=\frac{1}{4}\min\limits_{H_{A}}^{}[F(\rho_{AB,\theta})],
          \end{equation}
where $ F(\rho_{AB,\theta}) $ 
represents the QFI of the output state $\rho_{AB,\theta}=(U_{A}\otimes \mathcal{I}_{B})\rho_{AB}(U_{A}\otimes \mathcal{I}_{B})^{\dagger}  $ with respect to the phase.
 The minimization is performed over all Hamiltonians with the given non-degenerate spectrum $ \Lambda $ . This quantity, called Interferometric Power
(IP) of the state  $ \rho_{AB} $,  quantifies the minimum sensitivity in  interferometric
phase estimation.
It can be  proved that the IP has the same properties of the measures of non-classical
correlations, hence it can be used  as a discord-like quantity.

There is a simplified formula for the IP
of qubit-qudit systems, making it an easily
computable measure of non-classical correlations. In this case, the IP is obtained as \cite{Girolami2014,Lecture}

          \begin{equation}\label{IPqubit}
\text{IP}_{A}^{\Lambda}(\rho_{AB})=\chi_{min}(W_{AB}),
          \end{equation}
          that is,  the minimal eigenvalue of the $ 3\times 3 $-matrix $ M_{AB} $ whose elements are given by
          
                    \begin{equation}\label{Welement}
         (M_{AB})_{mn} =\frac{1}{2}\sum\limits_{i,j:p _{i}+p_{j}\neq 0}^{}\dfrac{(p_{i}-p_{j})^{2}}{p_{i}+p_{j}}\langle \psi_{i} |\sigma_{mA}\otimes \mathcal{I}_{B}|\psi_{j} \rangle_{AB} \langle \psi_{j} |\sigma_{nA}\otimes \mathcal{I}_{B}|\psi_{i} \rangle_{AB}
                    \end{equation}
where $ \sigma_{i} $'s denote the Pauli matrices.

\section{The Model   \label{Model}}
\subsection{Bipartite spin-boson model}
The first model that we  consider is
 two interacting two-level
systems, both coupled to an external reservoir
of bosonic field modes, i.e., two qubits coupled to
an environment of harmonic oscillators. 
The model can be mathematically described by the Hamiltonian of the free two-qubit system
$ H_{S} $, the  interaction Hamiltonian between the qubits and
the external bath $ H_{I} $, and the free Hamiltonian of the external
reservoir (bath) $ H_{B} $:
    \begin{equation}\label{HS}
H_{S}=\dfrac{\hbar \Omega_{1}}{2}\sigma_{z}^{1}+\dfrac{\hbar \Omega_{2}}{2}\sigma_{z}^{2}+\gamma \sigma_{z}^{1}\sigma_{z}^{2},
 \end{equation}
     \begin{equation}\label{HI}
 H_{I}=\sigma_{z}^{1}\otimes \sum\limits_{n=1}^{N}\lambda_{n}q_{n}+\sigma_{z}^{2}\otimes \sum\limits_{n=1}^{N}g_{n}q_{n},
  \end{equation}
      \begin{equation}\label{HSI}
  H_{B}=\sum\limits_{n=1}^{N}\hbar \omega_{n}a^{\dagger}_{n}a_{n}
   \end{equation}
where $ \Omega_{i} $ represents the characteristic frequency of $ i $th qubit, and $ \gamma $  is the coupling strength
    between the two spin qubits. Moreover, reminding the equivalence between the bosonic reservoir and the set of $ N $ quantized harmonic oscillators  characterized by frequencies $\{\omega_{n}\}  $ \cite{Scully},   we define
    $ \lambda_{n} $ and $ g_{n} $, respectively, as Spin 1 and Spin 2 coupling constants  to $ n $th
   oscillator in the environment. 
    A quantum
   Ohmic bath at zero temperature
    is  investigated and it is used subindexes $ 1,~2 $ or 12 referring to  different decoherence factors of both qubits. For instance, the decoherence factor appearing due to the interaction between qubit 1(2) and the environment is represented by $ \Gamma_{1(2)} $, while $ \Gamma_{12} $ denotes the  interaction between the composite system and the environment. On the other hand, defining $ J_{i}(\omega)=\gamma_{0i}/4\omega^{n}\Lambda^{n-1}\text{e}^{-\omega/\Lambda} $ as  the spectral density \cite{Leggett}  of the environment 
   associated to each spin of the system or the composite system, we absorb the coupling constants $ \lambda_{n}\equiv\lambda $ and $ g_{n}\equiv g $ in the dimensionless dissipative constants, i.e., $ \gamma_{01} \sim \lambda^{2},~ \gamma_{02} \sim g^{2}$, as well as  $ \gamma_{012} \sim \lambda g $,  and obtain the following forms for the decoherence factors  \cite{Fernando}:
   
      \begin{equation}\label{GAMMA1}
   \Gamma_{1}(t)=e^{-2\gamma_{01}\text{log}(1+\Lambda^{2}t^{2})},
    \end{equation}
           \begin{equation}\label{GAMMA2}
           \Gamma_{2}(t)=e^{-2\gamma_{02}\text{log}(1+\Lambda^{2}t^{2})},
            \end{equation}
       \begin{equation}\label{GAMMA3}
       \Gamma_{12}(t)=e^{-2\gamma_{012}\text{log}(1+\Lambda^{2}t^{2})},
        \end{equation}
where $ \Lambda $ denotes the environmental frequency cutoff.

The two qubits are initially prepared  in the following state

   \begin{equation}\label{INITIAL}
       \rho(0)=\frac{1-r}{4}\mathcal{I}+r|\vartheta\rangle \langle \vartheta|
        \end{equation}
where $ r \in (0,1] $, represents the mixing of the state and 
\begin{equation}\label{BELL}
       | \vartheta\rangle=\sqrt{1-p}|00\rangle+\sqrt{p}|11\rangle
        \end{equation}
in which $ p $ represents its degree of the entanglement. It is easy to see when $ p = 1/2 $, Eqs. (\ref{BELL}) is Bell state and Eq. (\ref{INITIAL}) defines the so-called Werner states playing a significant
role in quantum information processing. Moreover, a correspondence between the temperature T of the one-dimensional Heisenberg
 two-spin chain with a magnetic field B along the z axis and r of the Werner state
 has
been established \cite{Batle2005}.
\par
It can be proved that the evolved reduced density matrix is given by \cite{Fernando}

\begin{equation}\label{reduced1}
\rho_{1}(t)=\left(
\begin{array}{cccc}
\dfrac{1-r}{4}+r(1-p)& 0&0&r\sqrt{p(1-p)}e^{-i(\Omega_{1}+\Omega_{2})t}  \Gamma(t) \\
0 &\dfrac{1-r}{4}&0&0  \\
0 &0 &\dfrac{1-r}{4}&0  \\
r\sqrt{p(1-p)}e^{i(\Omega_{1}+\Omega_{2})t}  \Gamma(t)  & 0&0&\dfrac{1-r}{4}+rp   \\
\end{array} \right),
\end{equation}
where 
\begin{equation}\label{gamma}
  \Gamma(t)\equiv \Gamma_{1}(t)\Gamma_{2}(t)\Gamma_{12}^{2}(t).
  \end{equation}

  \subsection{Bipartite spin-spin model}
   One of the most  appropriate models in
the low-temperature regime is typically the spin environment. Particularly, experiments designed
to investigate the  macroscopic quantum coherence and decoherence require temperatures close to absolute zero for proper operation.
  We  consider the two-qubit system coupled to an external  environment composed of $ N $ spins, modelled by the Hamiltonians:
  
   \begin{equation}\label{HS2}
  H_{S}=\dfrac{\hbar \Omega_{1}}{2}\sigma_{z}^{1}+\dfrac{\hbar \Omega_{2}}{2}\sigma_{z}^{2}+\gamma \sigma_{z}^{1}\sigma_{z}^{2},
   \end{equation}
   \begin{equation}\label{HSI2}
       H_{E}=\sum\limits_{i=1}^{N} h_{i} \sigma_{xi},
        \end{equation}
       \begin{equation}\label{HI2}
   H_{I}=\sigma_{z}^{1}\otimes \sum\limits_{n=1}^{N}\varepsilon_{i}\sigma_{zi}+\sigma_{z}^{2}\otimes \sum\limits_{n=1}^{N}\lambda_{i}\sigma_{zi},
    \end{equation}
  where $ \xi_{i} $ ($ \lambda_{i} $)  denotes  the coupling
  between qubit 1 ( qubit 2)  and the spins of the environment, while $ h_{i} $ in the free Hamiltonian of the environment represents the tunneling matrix element
  for the ith-environmental spin.
  
  Preparing the initial state (\ref{INITIAL}), we find the following expression for  the reduced density matrix \cite{Fernando}
  
  \begin{equation}\label{reduced2}
  \rho_{2}(t)=\left(
  \begin{array}{cccc}
  \dfrac{1-r}{4}+r(1-p)& 0&0&r\sqrt{p(1-p)}e^{-i(\Omega_{1}+\Omega_{2})t} Q(t)  \\
  0 &\dfrac{1-r}{4}&0&0  \\
  0 &0 &\dfrac{1-r}{4}&0  \\
  r\sqrt{p(1-p)}e^{i(\Omega_{1}+\Omega_{2})t} Q(t)  & 0&0&\dfrac{1-r}{4}+rp   \\
  \end{array} \right),
  \end{equation}
 in which the decoherence factor $ Q(t) $ is given by
  \begin{equation}\label{Q}
  Q(t)=\prod\limits_{i=1}^{N}\bigg\lgroup  1-\big[\dfrac{2(\varepsilon_{i}+\lambda_{i})^{2}}{h_{i}^{2}+(\varepsilon_{i}+\lambda_{i})^{2}}\big] \text{sin}^{2}(t\sqrt{h_{i}^{2}+(\varepsilon_{i}+\lambda_{i})^{2}})\bigg\rgroup.
    \end{equation}

  \section{ Single parameter estimation}\label{Single}
  
  \subsection{Single parameter estimation in the presence of bosonic environment}
We consider the two-qubit system for \textit{independent} estimation of  parameters $ r $ and $ p $ encoded in the initial state. Changing the basis, we see that density matrix (\ref{reduced1}) can be written in the  block diagonal form
 \begin{equation}\label{D}
 \rho_{1}=\varrho_{1}\oplus \varrho_{2},
   \end{equation}
  where 
  \begin{equation}\label{D}
   \varrho_{1}=\left(
     \begin{array}{cc}
     \dfrac{1-r}{4}+r(1-p)& r\sqrt{p(1-p)}e^{-i(\Omega_{1}+\Omega_{2})t} \Gamma(t)  \\
     r\sqrt{p(1-p)}e^{i(\Omega_{1}+\Omega_{2})t} \Gamma(t)  & \dfrac{1-r}{4}+rp   \\
     \end{array} \right),
     \end{equation}
      and 
    \begin{equation}\label{D}
       \varrho_{2}=\left(
              \begin{array}{cc}
              \dfrac{1-r}{4}&0  \\
              0 & \dfrac{1-r}{4}  \\
              \end{array} \right).
         \end{equation}
Then, using the method introduced in \cite{J. Liu11,RanganiQIP4} (also see Sec. \ref{mohem} ) we compute the SLD operator, leading to following associated QFIs:

  \begin{equation}\label{Fofr}
 F_{i1}(r)= {\frac {3\,r-3-8\, \left( {\Gamma }^{2}-1 \right)  \left( p-1 \right) 
  p \left( 2\,r-1 \right) }{ \left( r-1 \right)  \bigg( r \bigg[ r
   \bigg( 16\, \left( {\Gamma }^{2}-1 \right)  \left( p-1 \right) p-3
   \bigg) +2 \bigg] +1 \bigg) }},
   \end{equation}
       \begin{equation}\label{Fofp}
  F_{i1}(p)=\frac {{r}^{2} \bigg[ \left( p-1 \right) p \left( 2\,e+r \right) ^{2
  }-{\Gamma }^{2} \bigg( e \left( 2\,p-1 \right) + \left( p-1 \right) r
   \bigg)  \bigg( e \left( 2\,p-1 \right) +pr \bigg)  \bigg] }{
   \left( p-1 \right) p \left( 2\,e+r \right)  \bigg( {e}^{2}+er+
   \left( {\Gamma }^{2}-1 \right)  \left( p-1 \right) p{r}^{2} \bigg)},
    \end{equation}
where
  \begin{equation}\label{e}
    e=\dfrac{1-r}{4}.
    \end{equation}
\begin{figure}[ht!]
                                 \subfigure[]{\includegraphics[width=6cm]{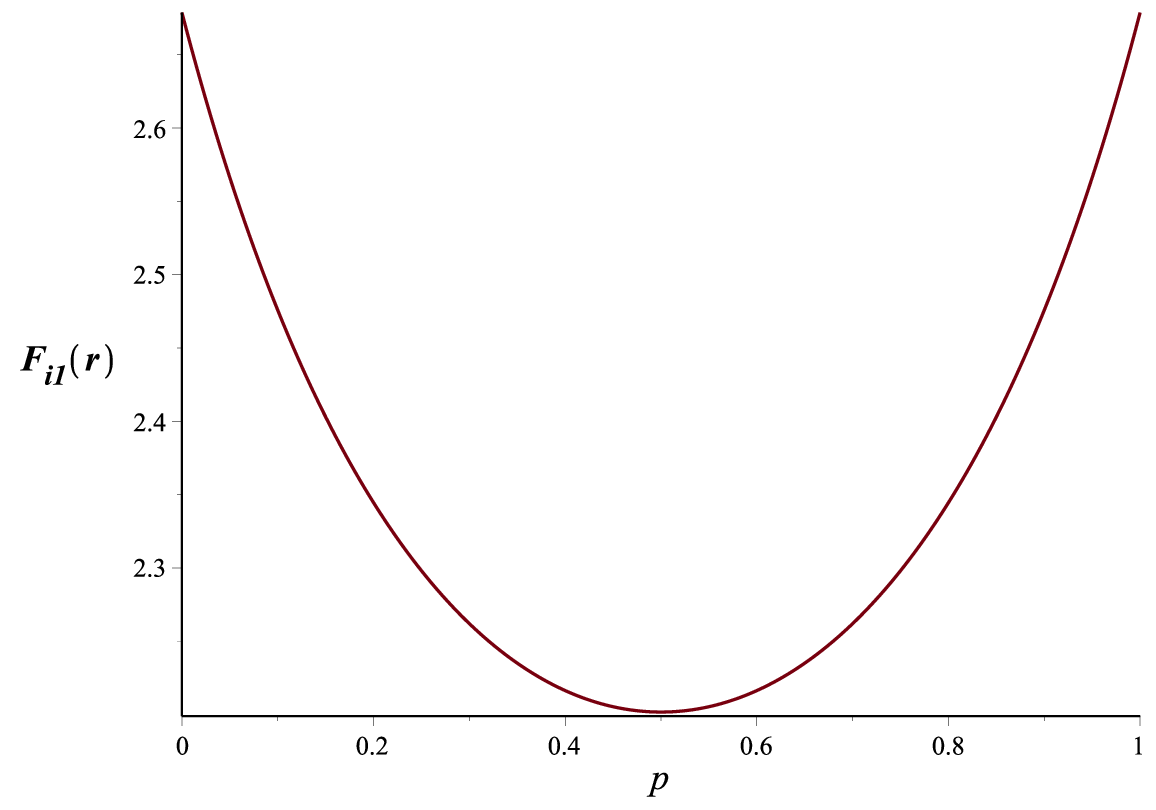}\label{Fi1rversusp}}
                                 \subfigure[]{\includegraphics[width=6cm]{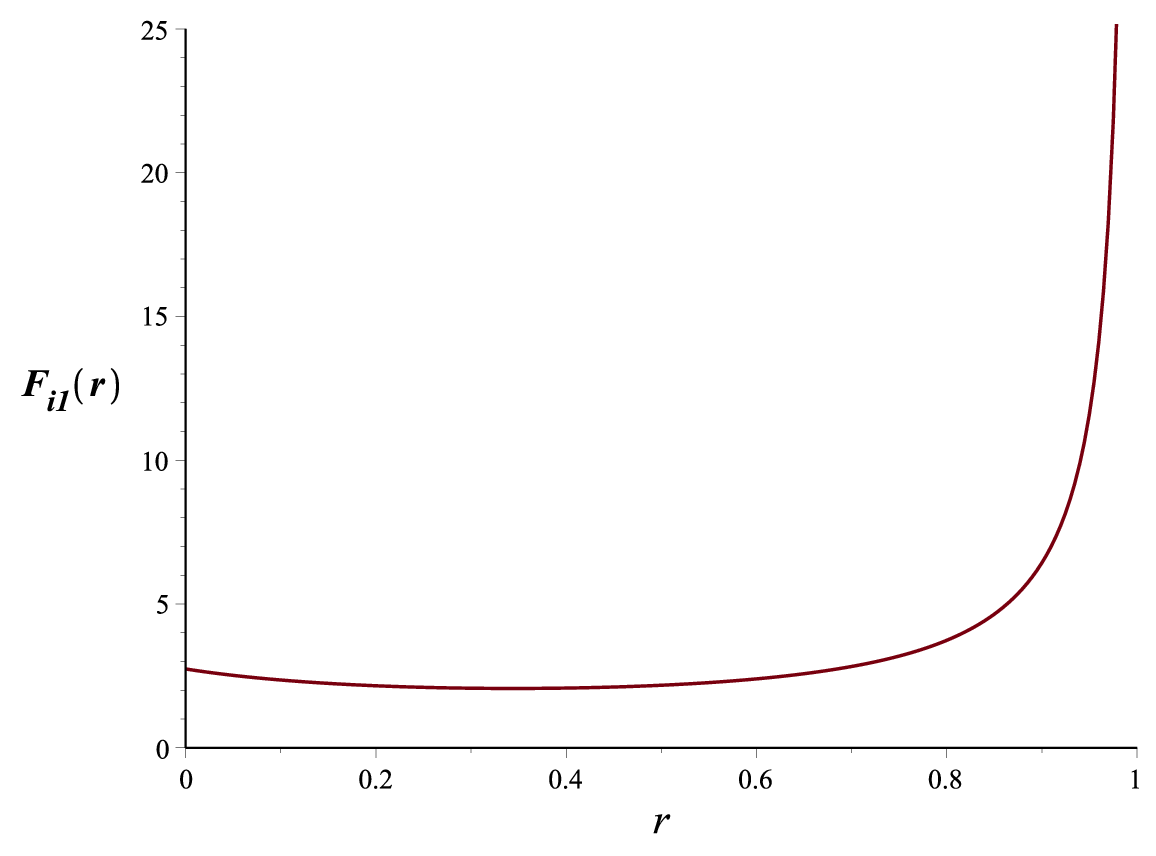}\label{Fi1rversusr}}
                                     \caption{\small   (a) QFI associated with
                                     independent estimation of the mixedness parameter as function of  $ p $ for $ r=0.6 $. (b) The same quantity versus $ r $ for $ p=0.4 $.}
                                                                      \label{Fi1rversuspr}
   \end{figure}
\begin{figure}[ht!]
                                 \subfigure[]{\includegraphics[width=6cm]{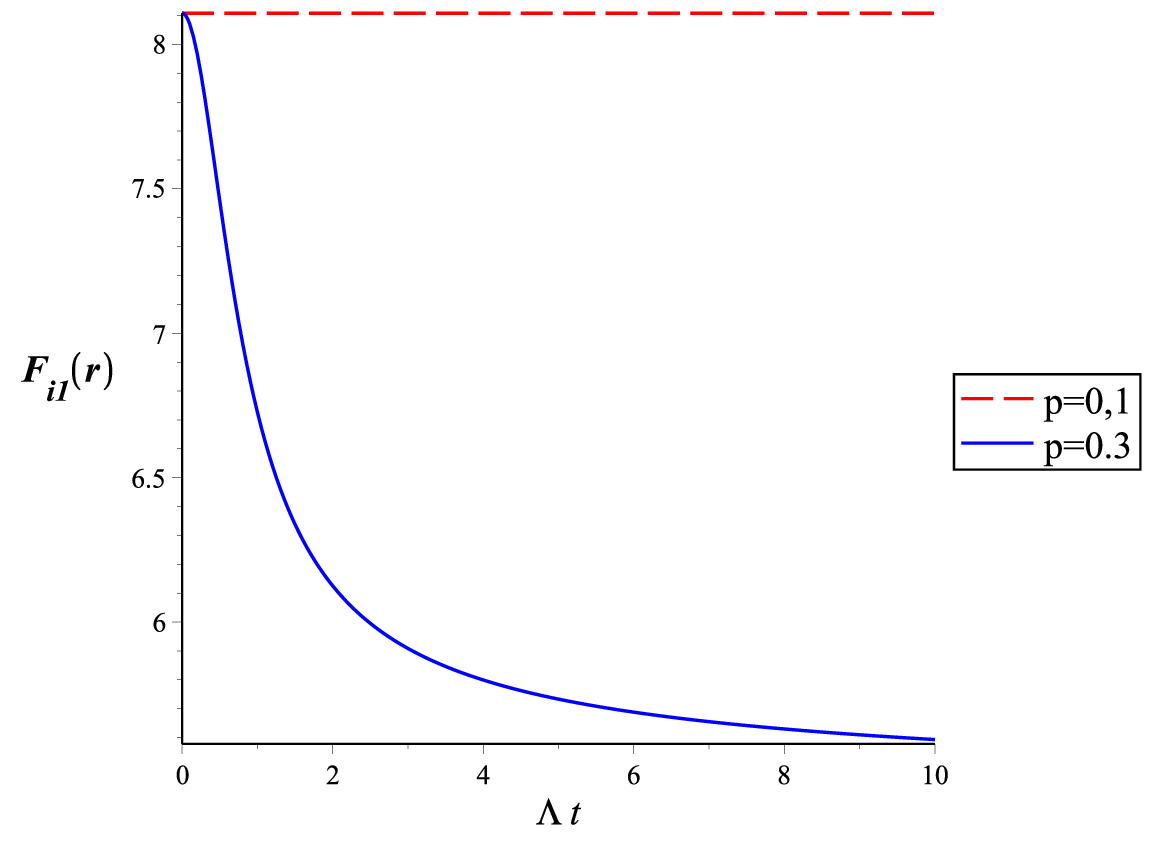}\label{Fi1rversust}}
                                 \subfigure[]{\includegraphics[width=6cm]{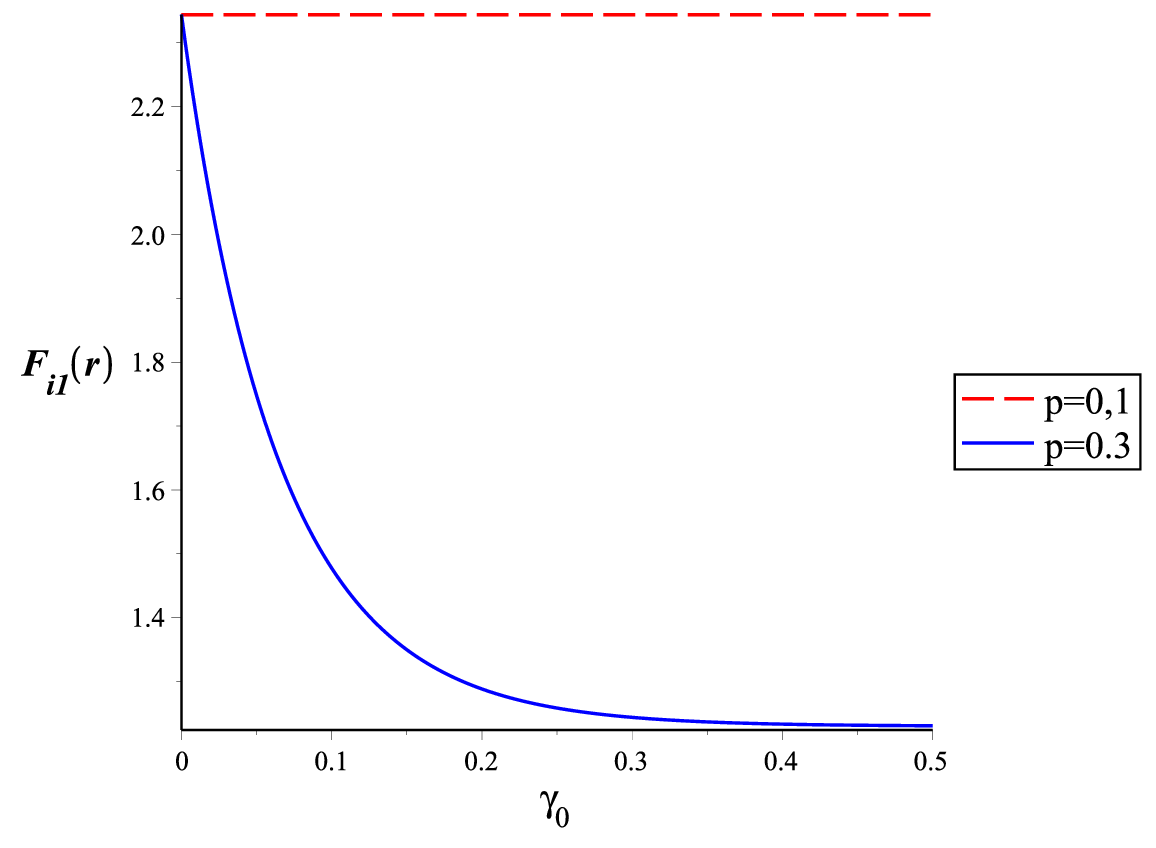}\label{Fi1rversusg0}}
                                     \caption{\small  (a) Dynamics of  QFI associated with
                                     independent estimation of  r 
                                     for $ \gamma_{0} =0.01$ and $ r=0.9 $. (b) The QFI
                                     as a function of the coupling constant for $ \Lambda t=1.2 $ and $ r=0.2 $.}
                                                                      \label{Fi1rversustg0}
   \end{figure}
   \begin{figure}[ht!]
                                    \subfigure[]{\includegraphics[width=6cm]{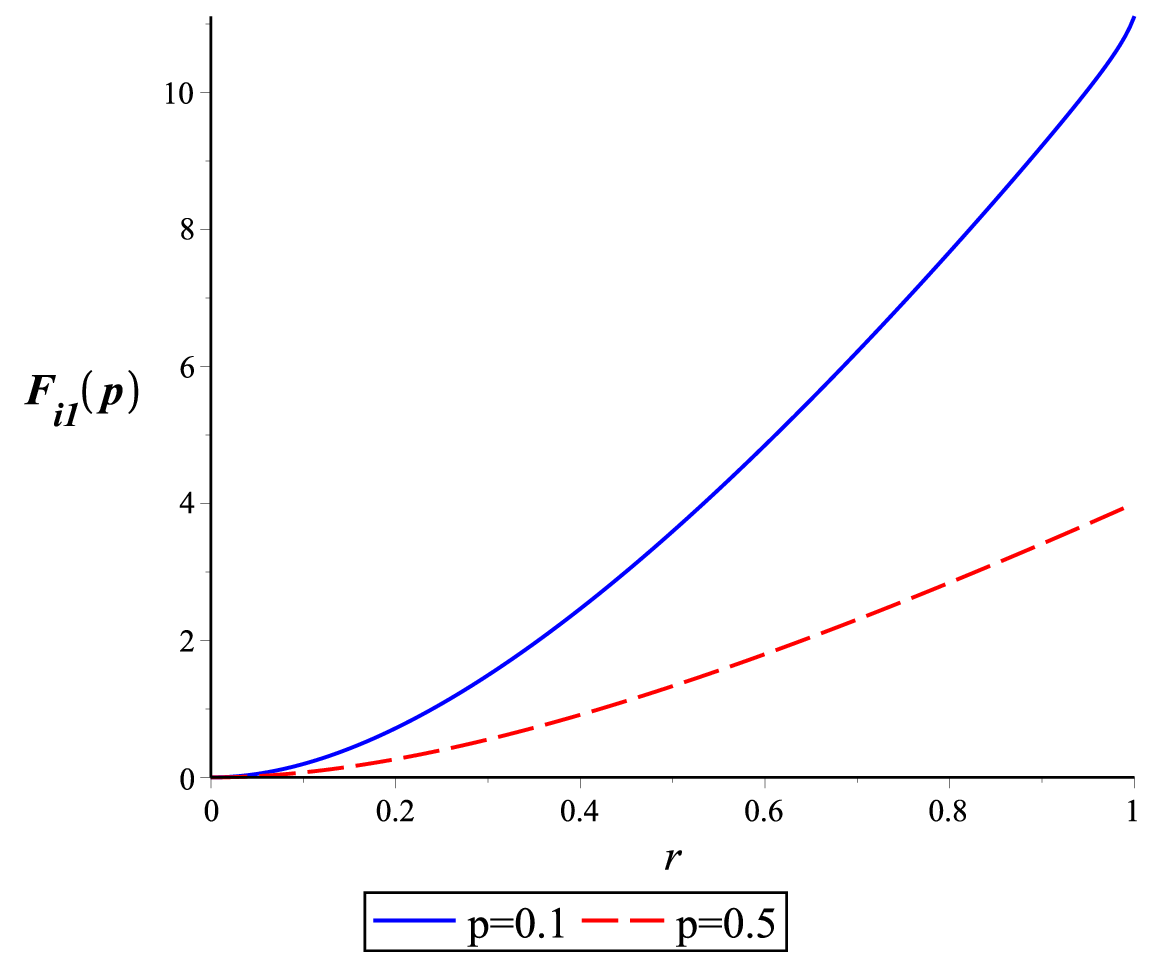}\label{Fi1pversusr}}
                                    \subfigure[]{\includegraphics[width=6cm]{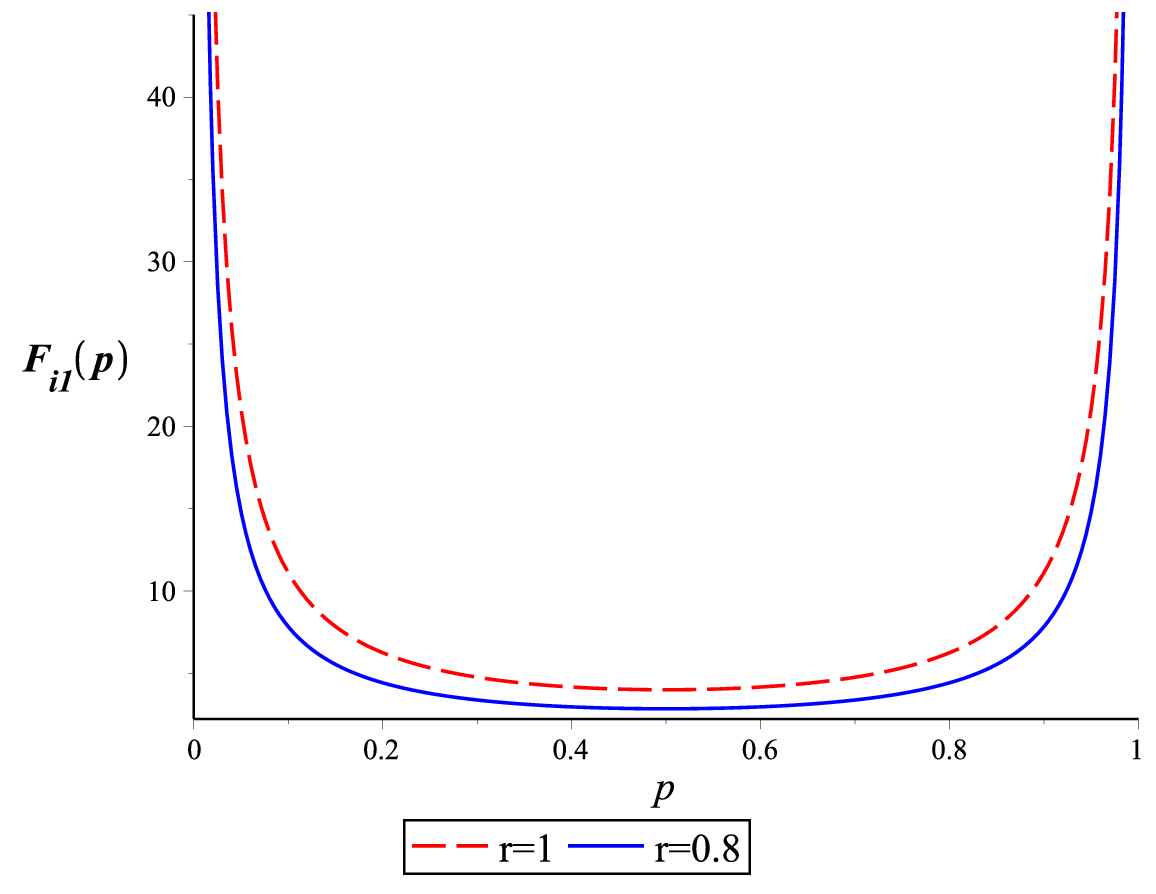}\label{Fi1pversusp}}
                                        \caption{\small (a) QFI associated with independent estimation of the degree of initial entanglement as a function of $ r $ for $ \Lambda t = 0.6$, $\gamma_{0}= 0.01 $. (b) The same quantity versus $ p $ for $ \Lambda t = 0.3$, $\gamma_{0}= 0.01 $ }
                                                                         \label{Fi1pversusrp}
      \end{figure}

We first focus on investigating the behaviour of $  F_{i1}(r)$, i.e., QFI associated with the mixedness parameter $ r $. Figure \ref{Fi1rversuspr}
  shows the variation of $  F_{i1}(r)$ versus entanglement and mixedness parameter. As seen in Fig. \ref{Fi1rversuspr}a, using more entangled initial states  leads to more loss of quantum information, extracted from the measurement in the process of estimation, and  consequently decrease of the QFI. The best estimation is obtained for ( $ p=0,1 $) (non-entangled initial state), and the minimum of the QFI occurs for $ p=1/2 $ at 
  which the entanglement of the initial state is maximized. This contradicts the well-known fact that entanglement usually enhances the estimation \cite{V. GiovannettiPRL,R. Demkowicz,Huelga1997,Kacprowicz2010,Chaves2013,Dinani2014,RanganiQIP3}.

  Figure  \ref{Fi1rversuspr}b  illustrates which values of $ r  $ are better estimated. Clearly, the best estimation is obtained for $ r=1 $, i.e.,  pure initial state $ \rho(0)=|\vartheta\rangle \langle \vartheta| $. Moreover, differentiating from  $ F_{i1}(r) $ with respect to $ r $ shows that when $ p=0,1$, for which the initial state entanglement is minimized, the minimum point of the QFI occurs at $ r=1/3 $.

  For  analysis of the QFI dynamics and its behaviour versus  the decoherence effect, we assume that the qubits are coupled to the bosonic
  ohmic environment by the same dimensionless coupling, i.e., $ \gamma_{1}=\gamma_{2}=\gamma_{12}=\gamma_{0} $. As seen in Fig. \ref{Fi1rversust}, the QFI dynamics exhibits a decreasing behaviour because of reduction in the precision of estimation. Similarly, a larger coupling constant leads to decrease in QFI originating from flow of information to the environment (see Fig. \ref{Fi1rversusg0}). On the other hand, there is a surprising  result using  a non-entangled initial state such that it causes  removal of these decoherence effects. In fact, the entanglement parameter $ p $ plays the role of a quantum key for turning off the decoherence effects in the process of estimating the mixedness parameter. 
In this condition, the QFI, given by the following expression,  is only dependent on  parameter r intending to estimate it;

 \begin{equation}\label{Fofpzero}
F_{i1}^{p=0,1}(r)={\frac {-3\,r+3}{ \left( r-1 \right) ^{2} \left( 3\,r+1 \right) }}
\end{equation}

Therefore, we obtain the \textit{QFI trapping} exhibiting asymptotic
behavior with the value given by Eq. (\ref{Fofpzero}). The primary reason is that
the bath cannot take the information during the interaction
with the qubits when  they have been prepared in a non-entangled initial state.

Figure \ref{Fi1pversusrp} shows the behaviour of the QFI associated with the independent estimation of the degree of initial state entanglement. As illustrated in Fig. \ref{Fi1pversusr}, the best estimation occurs when the qubits have been prepared in the  pure  state mathematically equivalent to  $ (r=1) $. In particular, One can \textit{freeze} the evolution of the QFI and protect it against the decoherece by using a pure initial state. Hence, the  corresponding QFI, i.e.,  its maximum value attainable in the process of  estimating the entanglement degree is given by

 \begin{equation}\label{Fofryek}
F_{i1}^{r=1}(p)=\dfrac{1}{p(1-p)}.
    \end{equation}
Clearly, it is independent of the decoherence effects. On the other hand, when the initial state is maximally entangled $ (p=1/2) $, the most inaccurate estimation  occurs (see Fig. \ref{Fi1pversusp}). Nevertheless, It is possible to detect the initial maximal entanglement by a trick; preparing the qubits in a maximally entangled state,  complete trapping of the QFI, associated with the entanglement degree is completely, is achieved. Particularly,  decoherece-free estimation again can be implemented according to the following expression of the QFI: 

\begin{equation}\label{Fofpnim}
F_{i1}^{p=1/2}(p)=\,{\frac {8{r}^{2}}{1+r}}
    \end{equation}

For other initial states, the QFI falls as time goes on or the coupling constant increases, similar to the behaviour of $F_{i1}(r)  $ shown in Fig. \ref{Fi1rversustg0} (blue line).

  \subsection{ Single parameter estimation in the presence of spin environment} 
         \begin{figure}[ht!]
                                            {\includegraphics[width=12cm]{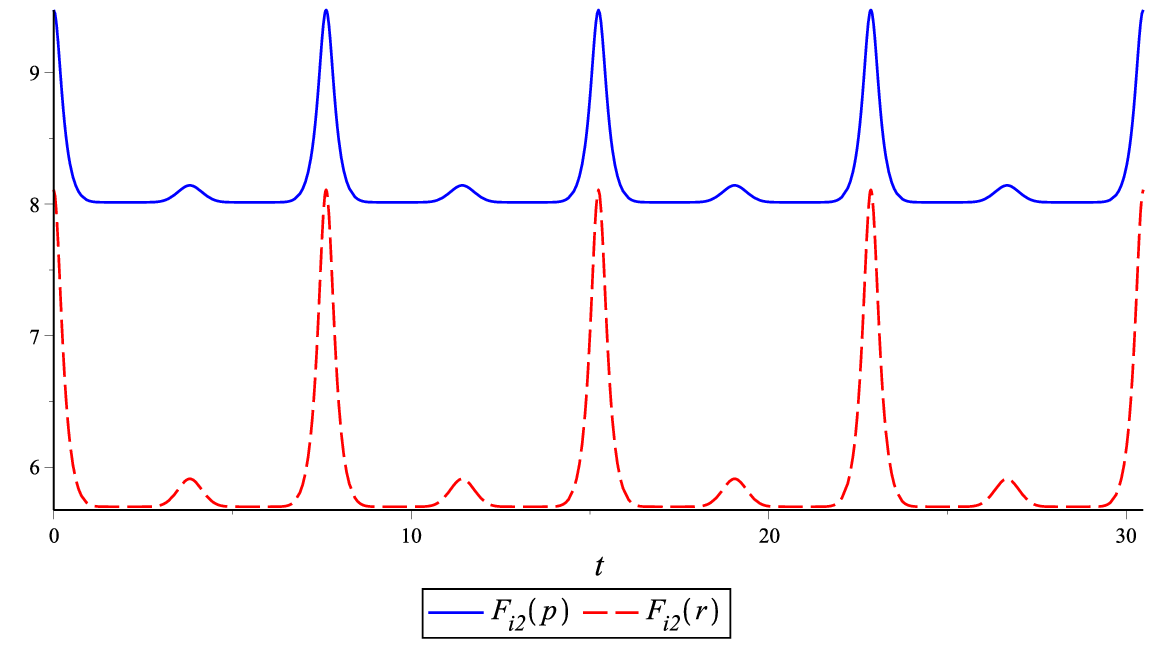}}
                                                \caption{\small  Dynamics of QFIs associated with independent estimations of the degree of initial entanglement and  the mixedness parameter  for 
                                                $ N = 5, p = 0.1, r = 0.9, h = 0.1 $, and  $ \lambda =0.2 $.
                                                \label{Fi2prversust} }
              \end{figure}
  \begin{figure}[ht!]
                                  \subfigure[]{\includegraphics[width=7cm]{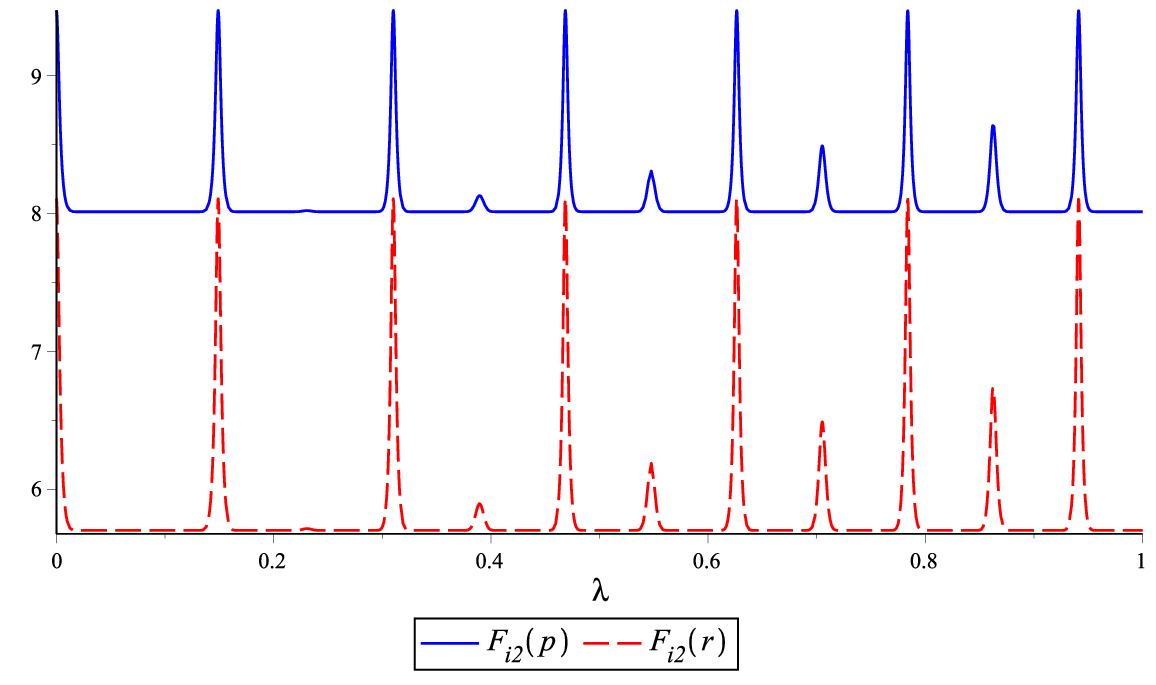}\label{Fi2prversusLambdaa}}
                                 \subfigure[]{\includegraphics[width=7cm]{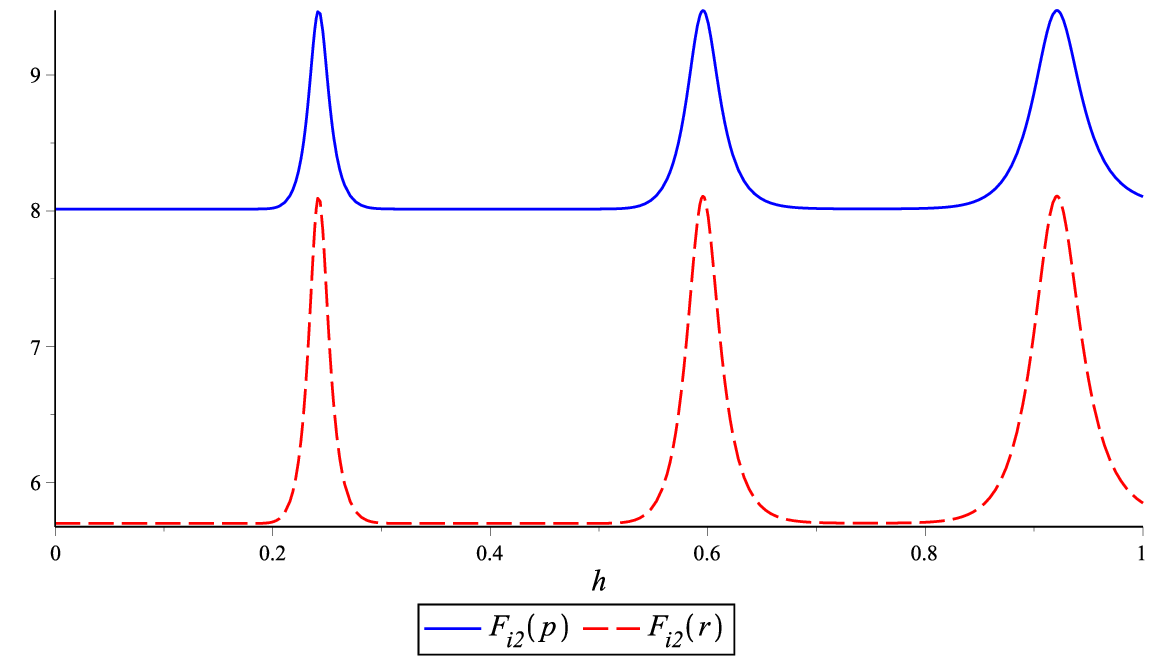}\label{Fi2prversush}}
                                       \caption{\small  (a) QFIs associated with independent estimations of initial parameters r and p  as
                                       functions of the coupling coefficient
                                        for 
                                 $ N = 20, p = 0.1, r =0 .9, h = 0.1, t = 10 $. (b) The same quantities as  functions of the tunneling parameter for $ N = 20, p = 0.1, r = 0.9, \lambda = 0.1, t = 10 $.
                                          \label{Fi2prversusLambdah} }
  \end{figure}
    \begin{figure}[ht!]
                                    \subfigure[]{\includegraphics[width=13cm]{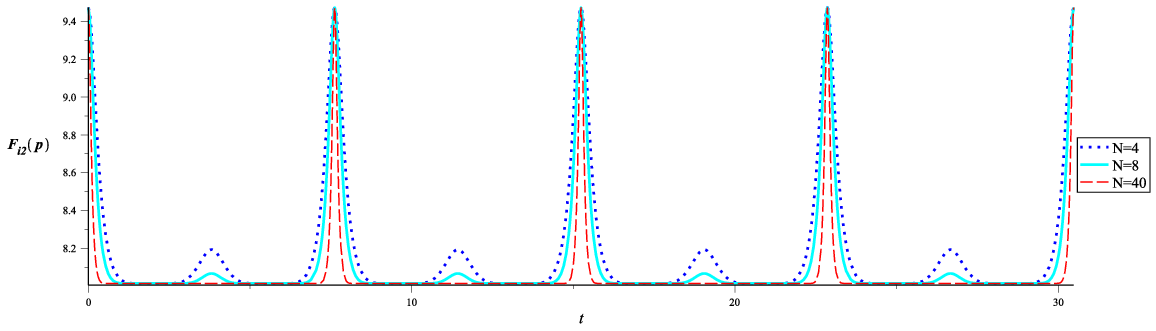}\label{Fi2prversusN1}}
                                   \subfigure[]{\includegraphics[width=13cm]{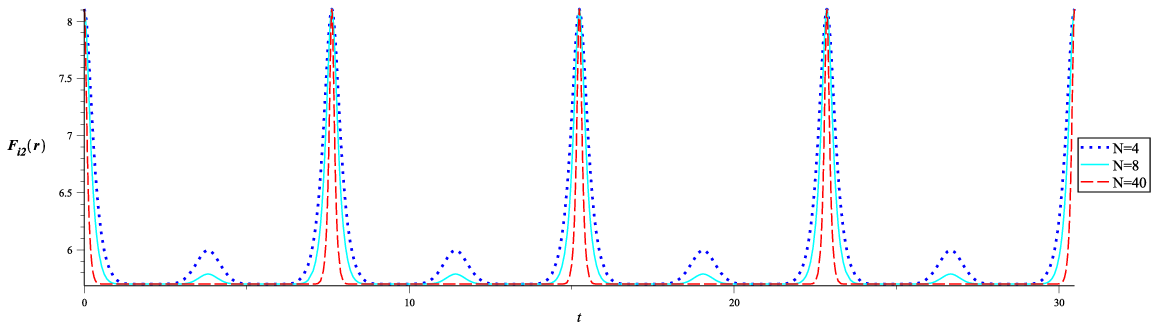}\label{Fi2prversusN2}}
                                         \caption{\small  QFIs associated with independent estimations of initial parameters $ p $ and $ r $  as
                                          functions of time  for 
                                   $ p = 0.1, r =0.9, h = 0.1, \lambda = 0.2 $ and different values of $ N $.
                                            \label{Fi2prversusN} }
    \end{figure}
  The  QFIs corresponding to parametrs $ r $ and $ p $ are given by replacing $ \Gamma $ with $ Q $ in Eqs. (\ref{Fofr}) and  (\ref{Fofp}), i.e.,
  
  \begin{equation}\label{Fofr2}
   F_{i2}(r)= {\frac {3\,r-3-8\, \left( {Q }^{2}-1 \right)  \left( p-1 \right) 
    p \left( 2\,r-1 \right) }{ \left( r-1 \right)  \bigg( r \bigg[ r
     \bigg( 16\, \left( {Q }^{2}-1 \right)  \left( p-1 \right) p-3
     \bigg) +2 \bigg] +1 \bigg) }},
     \end{equation}
         \begin{equation}\label{Fofp2}
    F_{i2}(p)=\frac {{r}^{2} \bigg[ \left( p-1 \right) p \left( 2\,e+r \right) ^{2
    }-{Q }^{2} \bigg( e \left( 2\,p-1 \right) + \left( p-1 \right) r
     \bigg)  \bigg( e \left( 2\,p-1 \right) +pr \bigg)  \bigg] }{
     \left( p-1 \right) p \left( 2\,e+r \right)  \bigg( {e}^{2}+er+
     \left( {Q }^{2}-1 \right)  \left( p-1 \right) p{r}^{2} \bigg)}.
      \end{equation}
  The results illustrated in Fig. \ref{Fi1rversuspr} as well as Fig. \ref{Fi1pversusrp} and ones obtained for the bosonic environment  in terms of the initial state, specially Eqs. (\ref{Fofpzero}-\ref{Fofpnim}) also holds for the spin environment. Moreover,  
   the QFIs \textit{synchronously} oscillate with time as they are
   suppressed to the minimum value and then rise, showing 
  the revivals of the information (see Fig. \ref{Fi2prversust} ).
  Within this regime,
  $ Q(t) $ presents oscillations describing a quasicoherent exchange
  of information between the qubits and the spin environment.
  This phenamena of collapse and revival  of the QFIs with time implies that the precision of estimation
  may enhance again during some time period. These oscillations can be regarded as evidence of the protection of information against the noise and the enhancement
  of coherence in the open quantum system, which originate from the reversed flow of information from the environment
  back to the quantum system.   Considering the coupling
  between each spin of the system and the external reservoir is
  equal, i.e., $ \varepsilon_{i}=\lambda_{i}\equiv \lambda $ and assuming $ h_{i}\equiv h $, we see that the period of oscillations is given by 
   \begin{equation}\label{T}
  T=\dfrac{2\pi}{\sqrt{h^{2}+4\lambda^{2}}} 
   \end{equation}
   and $ Q(t) $ becomes
       \begin{equation}\label{transformedQ}
      Q(t)=\bigg[ {\frac {8\, \bigg( \cos \left( t\sqrt {{h}^{2}+4\,{\lambda}^{2
      }} \right)  \bigg) ^{2}{\lambda}^{2}+{h}^{2}-4\,{\lambda}^{2}}{{h}^{2
      }+4\,{\lambda}^{2}}} \bigg] ^{N}.
        \end{equation}
Besides, we find that both QFIs simultaneously exhibit oscillatory behaviour such that the periods at which their collapses 
and revivals appear coincide. In other words, exchange of information about entanglement and mixedness parameter between the qubits and the spin environment occurs simultaneously. In particular, the maximum  points of the QFIs  coincide, which means that the estimations of those parameters,  
characterizing the initial state of the qubits,  are simultaneously optimized. 

\par
The variations of the QFIs with respect to the coupling coefficient and tunneling parameter are illustrated in Fig. \ref{Fi2prversusLambdah}. It is seen that the range of nonzero values of the QFIs as well as the points  at which the estimations of the initial parameters are optimized,  interestingly coincide. In addition, Fig. \ref{Fi2prversusLambdaa} shows that for both QFIs, investigated in terms of the coupling parameter, there are growing subordinate maxima  appearing between principal  maxima.

The QFIs are very sensitive to the size of the spin environment such that the
range of nonzero values of the QFIs, decreases as $ N $
increases (see Fig. \ref{Fi2prversusN}). Besides, as the
number of spins composing the environment is increased, the heights of the subordinate peaks
decrease while the principal peaks remain unchanged. Overall, apart from the number of spins composing the environment, the best estimations occur when the measurements are performed at instants 
$ t_{n}=nT,~(n=0,1,2,...) $ where $ T $ is given by  Eq. (\ref{T}).

\subsection{Lower bound on QFIs in terms of IP}
Recalling the definition of the IP in Sec. \ref{IPsec} and estimating the  imprinted phase by  a measurement at the output, it can be shown that \cite{Girolami2014} (the inverse of) non-classical correlations quantified by  IP upper bound the smallest possible variance of the estimator corresponding to interferometric phase estimations.
 In other words,  the existence of an
Hamiltonian and a measurement, such that the phase parameter  can be estimated with a
variance lower than a value determined by inverse  of IP, is  guaranteed.
It should be noted that in this  scenario 
perfect unitary evolution and ideal measurements are assumed, although we allow for noise in the
prepared input state before encoding the phase.
\par
Here our numerical calculation shows that the  QFIs of evolved state of the system, with respect  to independent estimations of parameters encoded in the qubits \textit{before} interaction with the environments, are lower bounded by the amount of the non-classical correlations of the noisy evolved state as quantified by the IP:

\begin{equation}\label{IPenquality}
\text{IP}(\rho(t)) \leq F_{i}(r),~ \text{IP}(\rho(t)) \leq F_{i}(p).
\end{equation}
Therefore, according to inequality (\ref{CRAMERsingle}),  the  quantum correlations of the evolved state,  measured
by the IP,  can guarantee an upper bound on the smallest possible
variance of  the initial parameter estimation performed  at each instant of time in the presence of bosonic or spin environment.
 Using Eq. (\ref{IPqubit}), one can obtain the IP for  quantum states  (\ref{reduced1}) and  (\ref{reduced2})
 \begin{equation}\label{IPasli}
\text{IP}(\rho(t))=\min \left( {\frac {{r}^{2} \left( e \left( 2-4\, \left( {g}^{2}-1
 \right)  \left( p-1 \right) p \right) + \left( {g}^{2}-1 \right) 
 \left( p-1 \right) pr \right) }{4\,{e}^{2}+2\,er+ \left( {g}^{2}-1
 \right)  \left( p-1 \right) p{r}^{2}}},-4\,{\frac {{g}^{2} \left( p-1
 \right) p{r}^{2}}{2\,e+r}} \right),
 \end{equation}
 where $ g $ stands for $ \Gamma $ and $ Q $  in the presence of bosonic and spin environment, respectively.

 \section{Flow of information}\label{flow}
 \subsection{Trace distance and memory effects}
 The trace norm defined by $\parallel \rho \parallel=\text{Tr}\sqrt{\rho^{\dagger}\rho}  $ leads to a 
 measure for the distance between two quantum states $ \rho^{1} $ and $ \rho^{2} $
 known as \textit{trace distance} \cite{Lecture} $ D(\rho^{1},\rho^{2})=\frac{1}{2}\parallel \rho^{1}-\rho^{2} \parallel $.
 It is  bounded
 by the inequality $ 0\leq  D(\rho^{1},\rho^{2}) \leq 1 $, where $ D(\rho^{1},\rho^{2})=0 $ if and only if
 $ \rho^{1}=\rho^{2} $, while it equals one if and only if the two states are
 orthogonal. We know that two density operators are said to be
 orthogonal provided that their \textit{supports}, defined as the subspaces spanned by their
 eigenstates with nonzero eigenvalue, are orthogonal. It can be shown that  trace distance $  D(\rho^{1},\rho^{2}) $ may be interpreted as the distinguishability  of   states $ \rho^{1} $ and $ \rho^{2} $. Moreover, any completely positive and trace preserving
 map (CPTP) $ \mathcal{E} $ is a contraction for the trace distance \cite{Breuer2016}, i.e., $ D\big(\mathcal{E}(\rho^{1}),\mathcal{E}(\rho^{2})\big) \leq D(\rho^{1},\rho^{2})$, for all quantum states $ \rho^{1,2} $. Since any  dynamical map $\mathcal{E}_{t}  $ describing  time evolution of an open quantum system is CPTP, the trace distance between
 the time-evolved quantum states can never be larger than the
 trace distance between the initial states. Hence, the
 dynamics generally diminishes the distinguishability of the states comparing it with the initial preparation. Certainly, this general
 fact does \textit{ not} imply that $ D\big(\rho^{1}(t),\rho^{2}(t)\big) $ where $ \rho^{1,2}(t)\equiv\mathcal{E}_{t}(\rho^{1,2}(0)) $ is a monotonically
 decreasing function of time \cite{Breuer2012}.

 According to BLP definition \cite{Breuer2009}, proposed by Breuer, Laine and Piilo (BLP),  
 a quantum evolution, mathematically described by a quantum
 dynamical maps $ \mathcal{E}_{t} $, is said to be Markovian when, for all pairs of initial
  states $ \rho^{1}(0) $ and $ \rho^{2}(0) $, the  trace
 distance $ D\big(\rho^{1}(t),\rho^{2}(t)\big) $ monotonically
 decreases  at all instants. Thus, 
quantum Markovian evolution  means a continuous
 loss of information from the open system to the environment.
 However, quantum memory effects arise if there is a temporal
 flow of information from the environment to the quantum system.
 The flowing back of information from the environment allows
 the earlier open system states to play a role in the later
 dynamics of the system, implying the emergence of
 the memory effects and non-Markovianity. 
 Hence, a quantum evolution  is called non-Markovian if there is an
 initial pair of states $ \rho^{1}(0) $ and $ \rho^{2}(0) $ such that the trace distance between the
 corresponding states $\rho^{1,2}(t)  $  is started to
 increase for a period of time: $ \dfrac{d}{dt}D\big(\rho^{1}(t),\rho^{2}(t)\big) >0 $. 
 
 \subsection{Non-Markovianity witnesses}
 \begin{figure}[ht!]
                                             {\includegraphics[width=12cm]{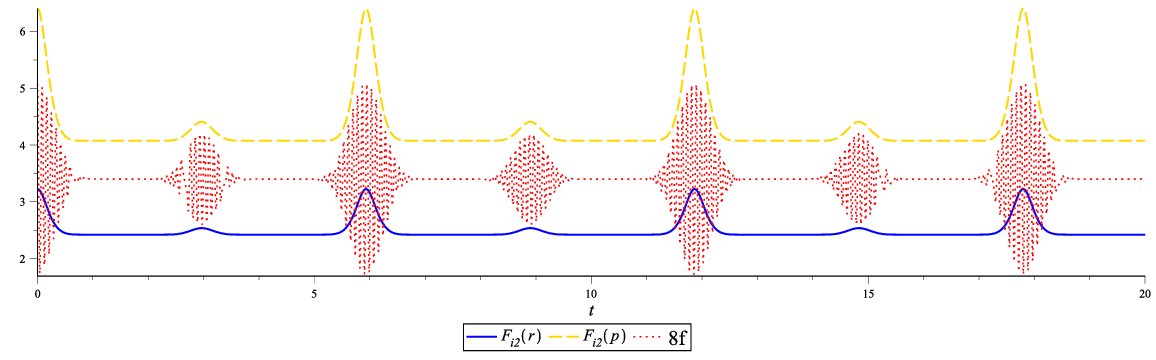}}
                                                 \caption{\small Comparison between the dynamics of QFIs and evolved fidelity    for 
                                                 $ N = 10, p = 0.1, r = 0.7, h = 0.1, \lambda = 0.26$, and  $ \Omega = 75 $.
                                                 \label{COMfid} }
               \end{figure}
               \begin{figure}[ht!]
                                                            {\includegraphics[width=12cm]{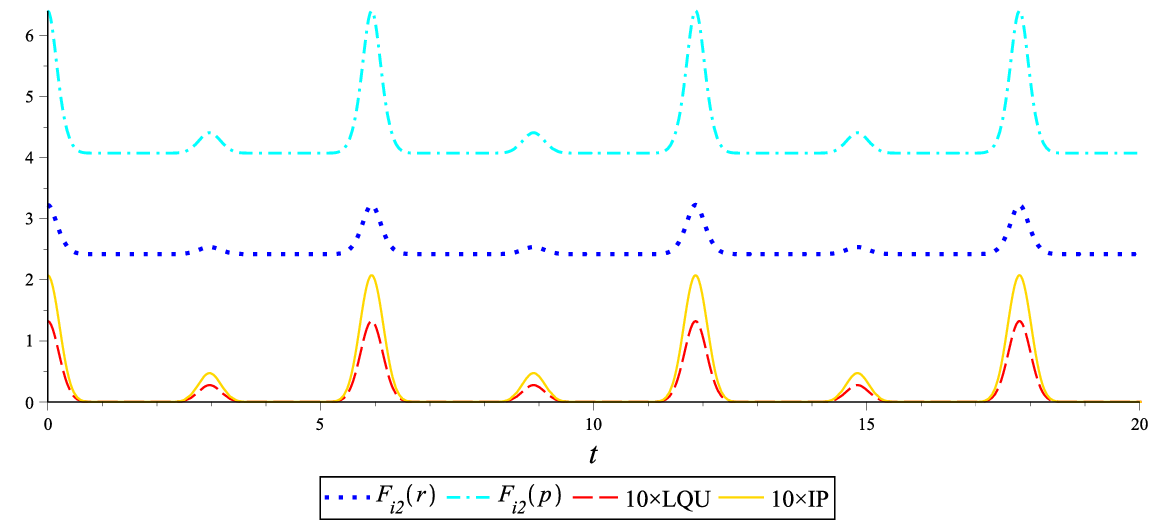}}
                                                                \caption{\small Comparison among the dynamics of QFIs, IP and LQU   for 
                                                                $ N = 10, p = 0.1, r = 0.7, h = 0.1$, and  $ \lambda = 0.26 $.
                                                                \label{FLQUIP} }
                              \end{figure}
 A witness of non-Markovianity is
 a quantity that can be used to detect the non-Markovian behaviour of the quantum system. The use of the QFI to witness non-Markovianity is originally due to Lu \textit{et al} in \cite{X2010}. These
 authors proved that the QFI is 
 monotonically decreasing under Markovian dynamics, 
therefore, if $ \dfrac{\partial F(\lambda)}{\partial t}> 0 $ for some period of time, the time evolution is non-Markovian \cite{RanganiAOP2,RanganiQIP3}.

Using the above criteria, here we unveil three important new non-Markovianity  witnesses, i.e., evolved fidelity, IP and LQU  for monitoring the memory effects in the second model.  Considering two quantum states $\rho^{1}  $ and $\rho^{2}  $ on the Hilbert space,  we can define the 
fidelity between those states as
$ f(\rho^{1},\rho^{2})=\text{Tr}\sqrt{\sqrt{\rho^{2}}\rho^{1}\sqrt{\rho^{2}}} $
measuring the degree of closeness of  quantum states $ \rho^{1} $ and $ \rho^{2} $. Inserting (\ref{INITIAL}) and (\ref{reduced2}) into this equation, we find the following expression for the \textit{evolved fidelity}:

\begin{equation}\label{EVOLfid}
f(\rho(0),\rho_{2}(t))=Q(t)r\sqrt {p \left( 1-p \right) }\cos \left( \Omega \right) +\dfrac{r}{4}+\frac{1}{4}
\end{equation}
where $ \Omega=\Omega_{1}+ \Omega_{2}$.

The variations of the QFIs and evolved fidelity versus time for $ N = 10, p = 0.1, r = 0.7, h = 0.1, \lambda = 0.26$, as well as $ \Omega = 75 $ are illustrated in Fig. \ref{COMfid}. 
It
is seen that, in  periods where the fidelity is constant, the QFIs also remain unchanged. On the other hand,  at instants when
 the \textit{collapse-revival} behaviour  of the fidelity is started, the enhancement of the QFIs  commences and consequently the non-Markovianity occurs because of information backflow from the environment to the system. In fact, in periods when the dynamics is non-Markovian, the fidelity exhibits \textit{growing} oscillatory behaviour, such that the optimum estimations are achieved at instants  when the fidelity is maximized.
 This result originates from the fact that we intend to estimate the initial parameters and the fidelity has been computed versus the initial state of the system. Moreover, the ranges of nonzero values of the QFIs coincide with the ranges at which the fidelity oscillates, such that in the periods when the QFIs increase (fall), the fidelity displays \textit{increasing} (\textit{falling})  oscillatory behaviour. Overall, in the period when the estimation precision is  enhanced (destroyed) and hence the dynamics is (non)-Markovian, the fidelity grows (falls) with oscillatory behaviour.
  Therefore, the evolved fidelity computed 
versus the initial state of the system can be used  as  an efficient  witness of
non-Markovianity. Interestingly, we see that the strength of fidelity collapse-revival behaviour is proportional to amplitude of QFIs oscillations, such that in the vicinity of subordinate maxima the amplitude of fidelity oscillations is also  low. Moreover, it should be noted that more larger (smaller)  frequency $ \Omega $ causes an increase (decrease) in the  frequency  with which the fidelity oscillates at each interval where the collapse-revival behaviour occurs.

Here we introduce the evolved state IP and LQU  as  significant witnesses of non-Markovianity (see Appendix \ref{APPB} for details of analytical computation of LQU). 
Figure \ref{FLQUIP} compares the  LQU and IP, which are the measures of non-classical
correlations of state (\ref{reduced2}),
 with the QFIs.
  This figure clearly illustrates that the three quantities
harmonically exhibit the same qualitative behaviour.
 When the evolution begins, the QFIs and hence precision of estimations decrease, leading to a loss of information about the encoded initial parameters. It  originates from decrease of  quantum correlations, quantified by IP and LQU, between the qubits which play an important role in 
 protecting the encoded information in the process of estimation.
  Later on,
all quantities increase, and consequently their time derivatives are positive
on this interval, i.e., $ \dfrac{d}{dt}\text{IP}>0,~ \dfrac{d}{dt}\text{LQU}>0 $, and $ \dfrac{d}{dt}\text{QFI}>0 $: this is the time interval when the non-Markovian behaviour
emerges, the system remembers past events, and  coherence is restored; hence the time derivatives $ \dfrac{d}{dt}\text{IP}$ and  $\dfrac{d}{dt}\text{LQU} $  are
positive over the non-Markovianity period, leading to introduce the LQU and the IP of the evolved state of the system
 as  witnesses of non-Markovianity in the presence of spin environment.

\section{Multiparameter estimation}\label{Multi}
\subsection{QFIM and QCRB} \label{mohem}
   \begin{figure}[ht!]
           \subfigure[]{\includegraphics[width=7cm]{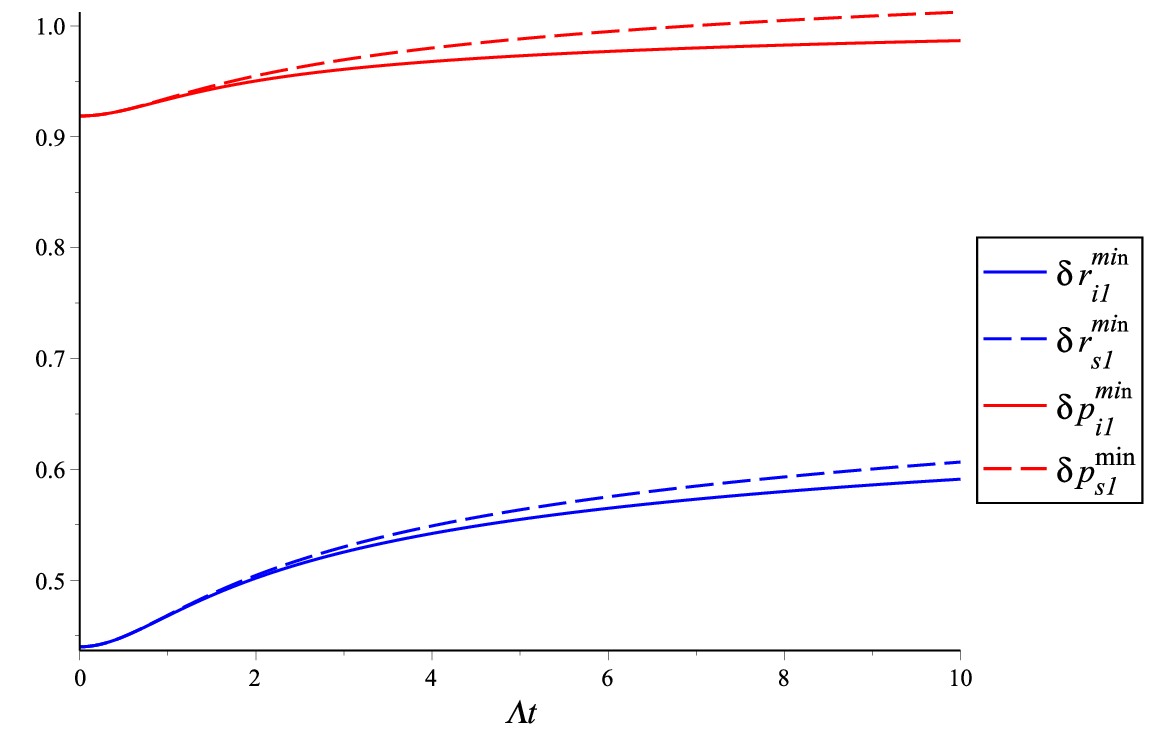}\label{QCRB1}}
           \subfigure[]{\includegraphics[width=7cm]{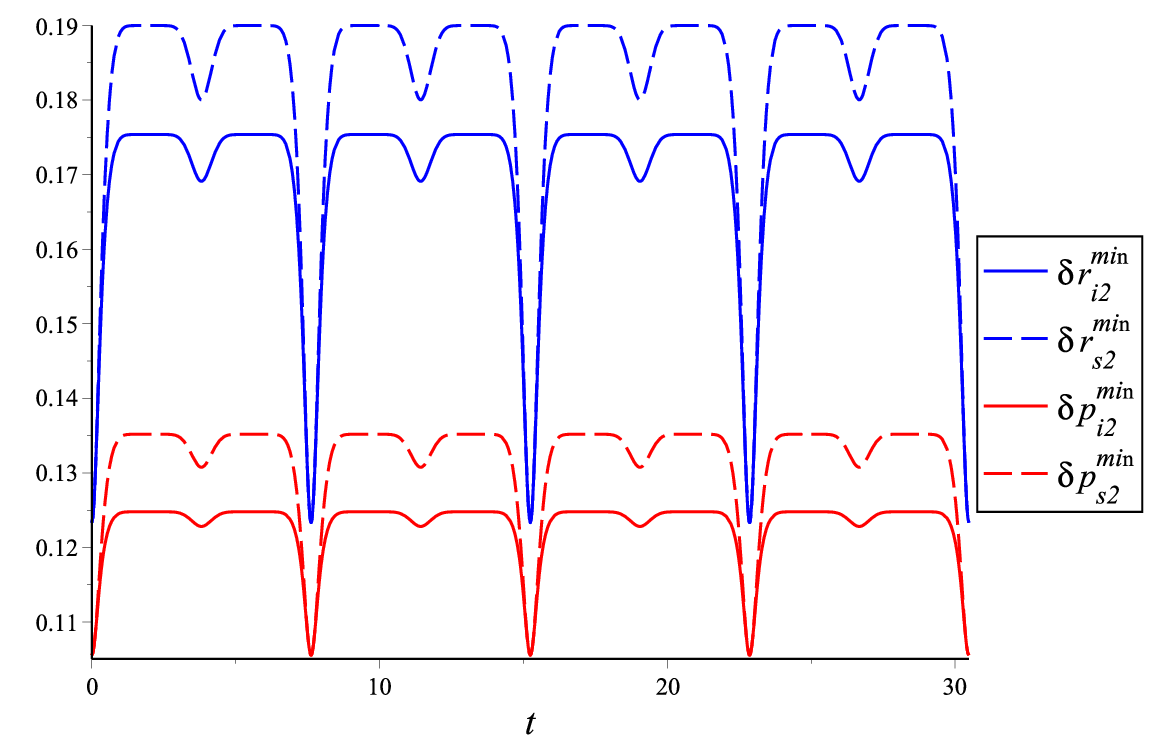}\label{QCRB2}}
                                         \caption{\small  (a) Time dependence of  QCRBs  for the first model with $ p = 0.3, \gamma_{0} = 0.01$ and  $r =0 .4 $. (b) The same quantities versus time for the second model with $ N = 5, p = 0.1, r =0.9, h = 0.1$ and  $ \lambda = 0.2 $.
                                            \label{QCRBplot} }
    \end{figure}
First, we should obtain SLDs $ L_{p} $ and  $ L_{r} $. After writing  block diagonal  state (\ref{reduced1}) or (\ref{reduced2})  in the form  $ \rho=\bigoplus^{n}_{i=1}\rho_{i}$, where $ \bigoplus $ denotes the direct sum, it  is clear that  the SLD operator may be  written as  $L_{\lambda_{j}}=\bigoplus^{n}_{i=1}L^{i}_{\lambda_{j}}$, where $L^{i}_{\lambda_{j}}$ represents the corresponding SLD operator with respect to parameter  $ \lambda_{j} $  for $\rho_{i}$.
    It has been shown that the SLD operator for the $i$th block is given by \cite{J. Liu11}
    \begin{equation}\label{SLDdiagonal}
      L^{i}_{\lambda_{j}}=\frac{1}{\mu_{i}}\left[\partial_{\lambda_{j}}\rho_{i}+\xi_{i}\rho^{-1}_{i} -\partial_{\lambda_{j}}\mu_{i} \right],
       \end{equation}
    where $\xi_{i}=2\mu_{i}\partial_{\lambda_{j}}\mu_{i}-\partial_{x}P_{i}/4$ in which $\mu_{i}=\text{Tr}\rho_{i}/2$ and $P_{i}=\text{Tr}\rho^{2}_{i}$. Note that $\xi_{i}$ vanishes if det$ \rho_{i}=0 $. 
    
    Following this method, we can construct the SLDs and obtain the QFIM. These results are presented  in Appendix \ref{A1}. In both models, using Eqs.  (\ref{CRAMERmulti}) and (\ref{InvQFIM}), it is seen that the variances of the \textit{simultaneous} estimations of  parameters are given by
 \begin{equation}\label{CRAMERmultiaslip}
        \delta p_{s}\geq (\textbf{F}^{-1})_{pp}={\frac {F_{{{\it rr}}}}{F_{{{\it pp}}}F_{{{
        \it rr}}}-{F_{{{\it pr}}}}^{2}}},
                  \end{equation}
\begin{equation}\label{CRAMERmultiaslir}
        \delta r_{s}\geq (\textbf{F}^{-1})_{rr}={
        \frac {F_{{{\it pp}}}}{F_{{{\it pp}}}F_{{{\it rr}}}-{F_{{{\it pr}}}}^{
        2}}}.  
                  \end{equation}
In our models, it can be proved that 
$ [L_{r},L_{p}]=0 $, hence these QCRBs can be achieved locally, i.e., a common measurement optimal from the point of
view of extracting information on both parameters $ r $ and $ p $ simultaneously, is realizable. Moreover,  we know that the corresponding uncertainty bounds for \textit{independent} estimations of the parameters are 
\begin{equation}\label{CRAMERsingleaslip}
        \delta p_{i}\geq \dfrac{1}{F_{pp}}\equiv\dfrac{1}{F_{i}(p)},
                  \end{equation}
\begin{equation}\label{CRAMERsingleaslir}
        \delta r_{i}\geq  \dfrac{1}{F_{rr}}\equiv\dfrac{1}{F_{i}(r)}.
                  \end{equation}

Because in both models the bounds are achievable, we plot the uncertainties for simultaneous estimations, $\delta p^{\text{min}}_{s}=(F^{-1})_{pp}  $ as well as  $\delta r^{\text{min}}_{s}=(F^{-1})_{rr}  $  
and compare them with the bounds for independent estimation cases, $\delta p^{\text{min}}_{i}=\dfrac{1}{F_{i}(p)} $ as well as  $\delta p^{\text{min}}_{i}=\dfrac{1}{F_{i}(r)} $. The scaling behaviour of these QCRBs as a function of time for the first and second model is plotted in Figs. \ref{QCRB1} and \ref{QCRB2}, respectively. We see that in the presence of bosonic or spin environment $\delta p_{i}\leq \delta p_{s}  $ and 
$\delta r_{i}\leq \delta r_{s}  $, hence the independent estimation of each parameter may lead to more accurate results than the simultaneous estimation.  Interestingly, as seen in Fig.  \ref{QCRB2}, even if the parameters are estimated simultaneously, the  oscillations of  QCRBs  are synchronous with those of the independent estimations.

\subsection{Role of coherence and purity in multiparameter estimation}
   \begin{figure}[ht!]
                                    \subfigure[]{\includegraphics[width=7cm]{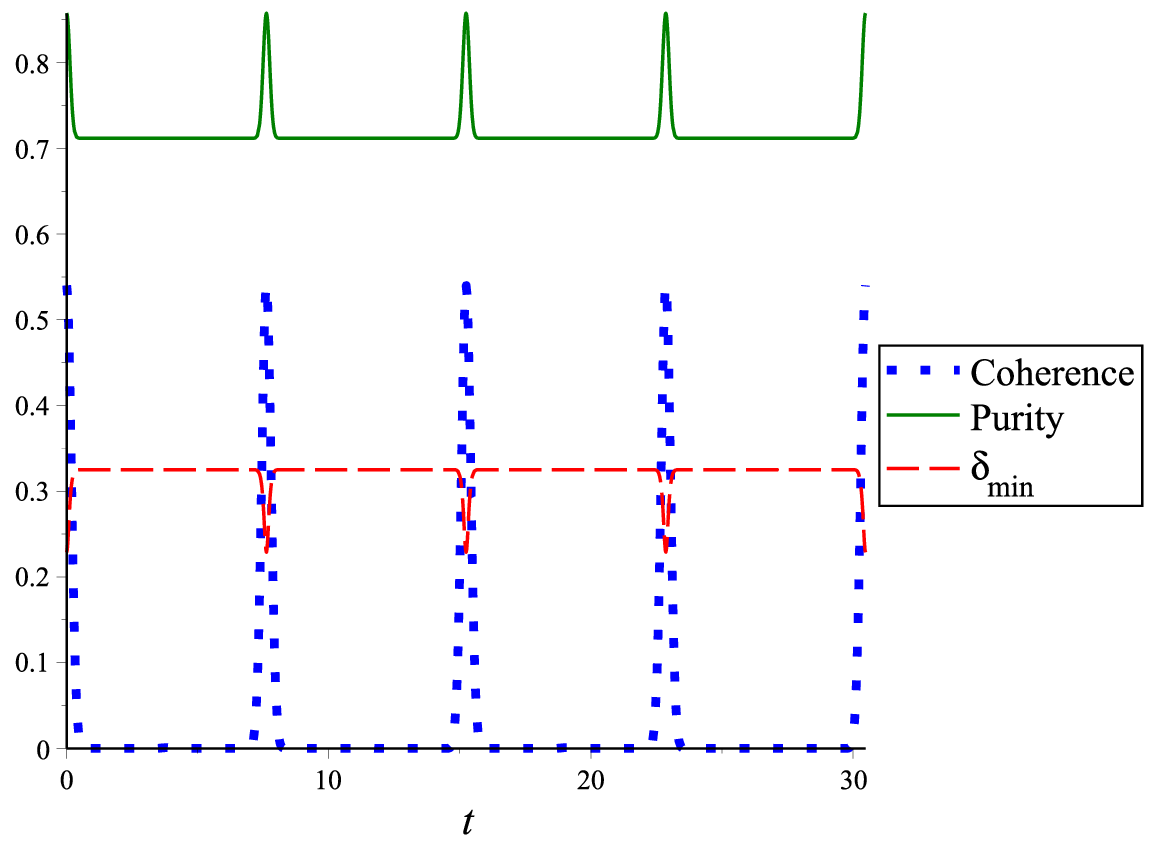}\label{Cpdel1}}
                                   \subfigure[]{\includegraphics[width=7cm]{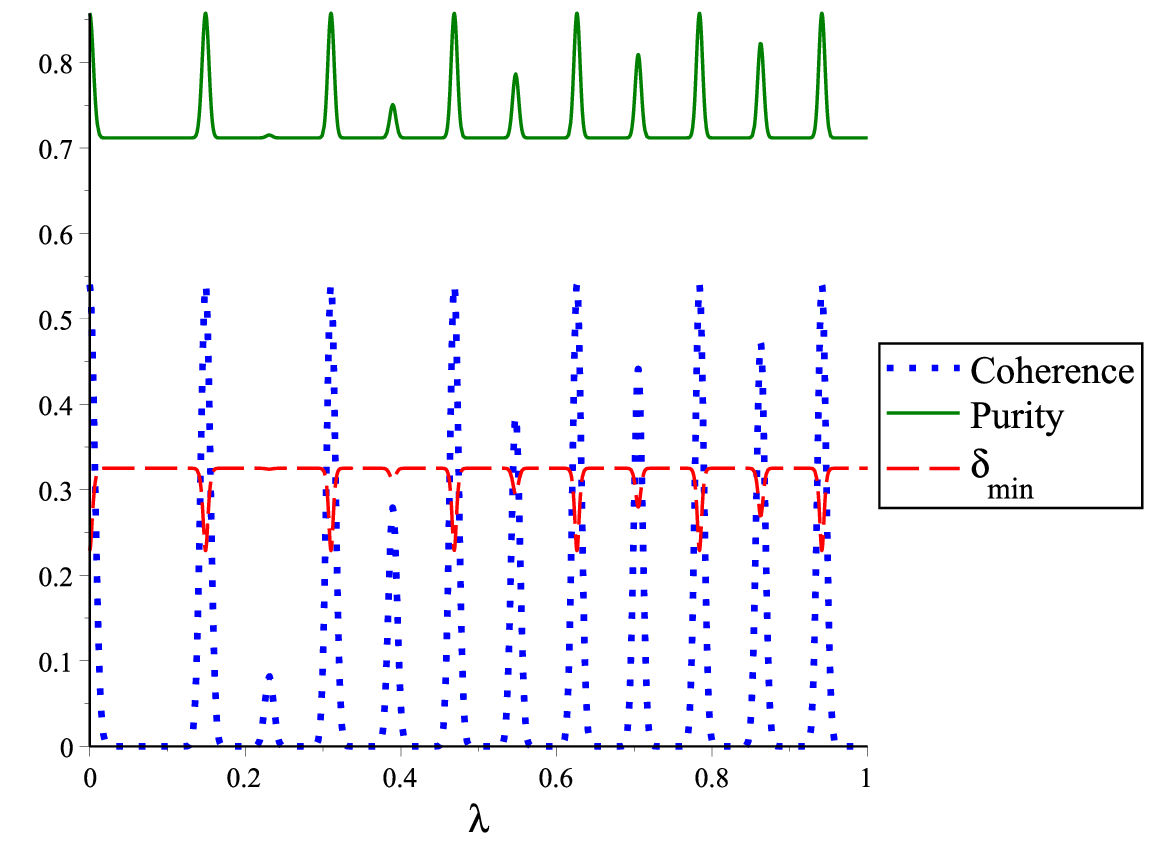}\label{Cpdel2}}
                                         \caption{\small  (a) Time dependence of  the coherence, purity and total variance  for the second model with $ N = 50, p = 0.1, r = 0.9, h = 0.1$ and  $\lambda = 0.2 $. (b) The same quantities versus $ \lambda $  for  $ N = 20, p = 0.1, r = 0.9, h = 0.1$ and  $ t=10 $.
                                            \label{Cpdel} }
    \end{figure}
Introducing the intuitive $ l_{1} $-norm measure of quantum coherence \cite{Baumgratz2014}, which quantifies the coherence in the reference basis through the off-diagonal elements of density matrix, $ C_{l_{1}}\left(\rho \right)=\sum_{i,j\atop i\ne j}|\rho_{ij}| $, we obtain the following expression for quantum coherence
  of  the evolved state of the system:
\begin{equation}\label{coherence}
 C_{l_{1}}\left(\rho \right(t))=2r\sqrt{p(1-p)}|g|.
 \end{equation}
 Moreover, the evolved state \textit{purity} $ P(\rho)=\text{Tr}\big(\rho^{2}\big) $,
  a particular measure of the   quantum state noisiness, is given by
  \begin{equation}\label{Purity}
   P(\rho(t))=-2\,{g}^{2}{r}^{2} p \left( p-1 \right) +\frac{1}{4}\, \left( 8\,{p}^{2}-8
   \,p+3 \right) {r}^{2}+\frac{1}{4}
   \end{equation}
  where as before $ g $ stands for $ \Gamma $ and  $ Q $, discussing the first and second model, respectively. While  the
 purity of a pure state is equal to one, the purity of a mixed one is strictly
 less than 1. Besides, the minimum value of the purity  is bounded by the inverse
 of the dimension of the system Hilbert space.
 \par
 We investigate the relationship among  coherence as well as purity of the quantum state and lower  bound on the total variance of all the parameters that should be estimated, i.e.,
  \begin{equation}
\delta_{\text{min}}=\text{Tr}(\textbf{F}^{-1})={\frac {F_{{{\it rr}}}+F_{{{\it pp}}}}{F_{{{\it pp}}}F_{{{\it rr}}}-{F
 _{{{\it pr}}}}^{2}}}.
  \end{equation}

 Figure \ref{Cpdel1} shows that these quantities synchronously oscillate with time such that at instants when purity and coherence of the reduced density matrix are constant, the total variance does not
 change with time. With increase (decrease) in coherence or purity, the total variance  is started to
 decrease (increase) simultaneously in the sense that the minimum values of the total variance are obtained when  the coherence and purity 
 are maximized. Therefore, we expect that the optimized simultaneous estimation of the initial parameters occurs at the  instants when the quantum state coherence is maximized and the purity tends  toward one, equivalent to pure quantum state. In other words, coherence and purity of the evolved state of the probes are two key elements for realizing optimum multiparameter estimation. Another aspect of these close relationship is shown in Fig. \ref{Cpdel2} illustrated the variation of those quantities versus coupling coefficient $ \lambda $. We see that all those quantities exhibit growing subordinate maxima  appearing between principal  maxima.

  \section{Summary and conclusions \label{conclusion}}
  \par To summarize, we investigated the independent and simultaneous  parameter estimation problem for two interacting two-level systems, both coupled to  external bosonic and spin environments. 
   We found that the entanglement parameter  plays the role of a quantum key for turning
  off the decoherence effects in the process of estimating the mixedness parameter. Besides, one can freeze the evolution of the QFI, corresponding to estimation of the  entanglement parameter,  and protect it against the decoherece by using a pure initial state.
   On the other hand, the quantum correlations measured
    by the IP, are a sufficient resource to guarantee an upper bound on the smallest possible variance of the initial parameter estimation. 
   It was also found while the memory effects are not appeared in the presence of bosonic environment, the dynamics of the system in the presence spin environment is non-Markovian.  
   We unveiled the evolved state  IP and LQU and fidelity as new witnesses of non-Markovianity, 
  that can be used to detect the backflow of information from the
  spin environment to the system.
  Besides,  we computed analytically the   two qubit QFIM for multiparameter estimation and investigated the corresponding  QCRBs
 in both single and multiparameter estimations. In particular, the relationships among quantum coherence, purity and multiprameter estimation were discussed.

\section*{Acknowledgements}

 H. R. J. wishes to acknowledge the financial support of
 the MSRT of Iran and Jahrom University.

\appendix

\section{SLD and QFIM elements}\label{A1}

\par  
Using Eq. (\ref{SLDdiagonal}), we obtain the following expressions for the SLDs:

\begin{equation}\label{Lp}
L_{p}=\left(
\begin{array}{cccc}
{\frac {r}{2\,e+r} \left( {\frac { \left( {g }^{2}-1 \right) 
 \left( 2\,p-1 \right) r \left( pr+e \right) }{{e}^{2}+er+ \left( {
g }^{2}-1 \right)  \left( p-1 \right) p{r}^{2}}}-2 \right) }
& 0&0&-{\frac {  g   \,e \left( 2\,p-1 \right) r \left( e+
r \right) {{\rm e}^{-it\Omega}}}{\sqrt {- \left( p-1 \right) p}
 \left( 2\,e+r \right)  \left( {e}^{2}+er+ \left( {g }^{2}-1
 \right)  \left( p-1 \right) p{r}^{2} \right) }}
  \\
0 &0&0&0  \\
0 &0 &0&0  \\
-{\frac {  g   \,e \left( 2\,p-1 \right) r \left( e+
r \right) {{\rm e}^{it\Omega}}}{\sqrt {- \left( p-1 \right) p}
 \left( 2\,e+r \right)  \left( {e}^{2}+er+ \left( {g }^{2}-1
 \right)  \left( p-1 \right) p{r}^{2} \right) }} & 0&0&{\frac {2}{2\,e+r} \left( {\frac { \left( {g }^{2}-1 \right) 
  \left( 2\,p-1 \right) {r}^{2} \left( -pr+e+r \right) }{2\,{e}^{2}+2\,
 er+2\, \left( {g }^{2}-1 \right)  \left( p-1 \right) p{r}^{2}}}+r
  \right) }
  \\
\end{array} \right),
\end{equation}
\begin{equation}\label{Lr}
L_{r}=\left(
\begin{array}{cccc}
{\frac { \left( 16\,{g}^{2}{p}^{2}-16\,{g}^{2}p-16\,{p}^{2
}+16\,p-3 \right) r-4\,p+3}{1+ \left( 16\,{g}^{2}{p}^{2}-16\,{
g }^{2}p-16\,{p}^{2}+16\,p-3 \right) {r}^{2}+2\,r}}
& 0&0&{\frac { 4 g \sqrt {-
 \left( p-1 \right) p}  \,{{\rm e}^{-it\Omega}}}{1+ \left( 16\,{g }^{2}{p}^{2}-16\,{
g }^{2}p-16\,{p}^{2}+16\,p-3 \right) {r}^{2}+2\,r}}
  \\
0 &0&0&0  \\
0 &0 &0&0  \\
{\frac { 4 g \sqrt {-
 \left( p-1 \right) p}   \,{{\rm e}^{it\Omega}}}{1+ \left( 16\,{g }^{2}{p}^{2}-16\,{
g }^{2}p-16\,{p}^{2}+16\,p-3 \right) {r}^{2}+2\,r}}
 & 0&0&{\frac { \left( 16\,{g }^{2}{p}^{2}-16\,{g }^{2}p-16\,{p}^{2
 }+16\,p-3 \right) r+4\,p-1}{1+ \left( 16\,{g }^{2}{p}^{2}-16\,{
 g }^{2}p-16\,{p}^{2}+16\,p-3 \right) {r}^{2}+2\,r}}
  \\
\end{array} \right),
\end{equation}
where $ \Omega=\Omega_{1}+\Omega_{2} $ and $ g $ stands for $ \Gamma $ and $ Q $ in the presence of bosonic and spin environment, respectively.
Moreover,  we find that the elements of the QFIM
\begin{equation}\label{Lr}
\textbf{F}=\left(
\begin{array}{cc}
F_{pp}
&F_{pr}
  \\
F_{rp} &F_{rr}  \\
\end{array} \right),
\end{equation}
 corresponding to simultaneous estimation of $ p $ and $ r $, are given by:
    \begin{equation}\label{F1rp}
F_{pr}=F_{rp}=\,{\frac {-4 \left( {g }^{2}-1 \right)  \left( 2\,p-1 \right) r}{
r \left( r \bigg[ 16\, \left( {g }^{2}-1 \right)  \left( p-1
 \right) p-3 \bigg] +2 \right) +1}},
 \end{equation}
     \begin{equation}\label{F1rr}
F_{rr}\equiv F_{i}(r),
  \end{equation}
      \begin{equation}\label{F1pp}
 F_{pp}\equiv F_{i}(p).
   \end{equation}
 Besides, the inverse matrix of the QFIM is given by:
\begin{equation}\label{InvQFIM}
\textbf{F}^{-1}=\left( \begin {array}{cc} {\frac {F_{{{\it rr}}}}{F_{{{\it pp}}}F_{{{
\it rr}}}-{F_{{{\it pr}}}}^{2}}}&-{\frac {F_{{{\it pr}}}}{F_{{{\it pp}
}}F_{{{\it rr}}}-{F_{{{\it pr}}}}^{2}}}\\ \noalign{\medskip}-{\frac {F
_{{{\it rp}}}}{F_{{{\it pp}}}F_{{{\it rr}}}-{F_{{{\it pr}}}}^{2}}}&{
\frac {F_{{{\it pp}}}}{F_{{{\it pp}}}F_{{{\it rr}}}-{F_{{{\it pr}}}}^{
2}}}\end {array} \right).
\end{equation}
 
\section{Computation of LQU}\label{APPB}
\par
Considering  a bipartite quantum system prepared in the state $ \rho=\rho_{AB} $, we  suppose that $ O^{\varLambda}\equiv O^{\varLambda}_{A} \otimes \mathcal{I} _{B}  $ denotes a local observable, in which $  O^{\varLambda}_{A}$ represents a Hermitian operator on
$ A $ with non-degenerate spectrum $ \varLambda $.
The LQU with respect to subsystem $ A $ is defined as follows \cite{Girolami2013}
          \begin{equation}
\text{LQU}^{\varLambda}_{A}=\min_{O^{\varLambda}}I(\rho,O^{\varLambda}).
          \end{equation}
where  $ I(\rho,O^{\varLambda})=-\dfrac{1}{2} \text{Tr}\{[\sqrt{\rho},O^{\varLambda}]^{2}\} $   denotes the \textit{skew information}. Moreover, the minimization is performed over  all local observables of $ A $ with non-degenerate spectrum $ \varLambda $.

However, for a qubit-qudit system,  the choice
of the spectrum $ \varLambda $ does not affect the quantification of non-classical correlations, therefore 
we can drop the $ \varLambda$ superscript from here onwards. Besides, for qubit-qudit systems,it is possible to write the LQU in the following form: 
          \begin{equation}
\text{LQU}_{A}=1-\lambda_{max}\big(W_{AB}\big),
          \end{equation}
in which $\lambda_{max}\big(W_{AB}\big)  $ represents  the maximum eigenvalue of the $ 3\times 3 $ symmetric matrix $ W $ with
elements given by:
          \begin{equation}
(W_{AB})_{ij}=\text{Tr}\big[\sqrt{\rho_{AB}}(\sigma_{iA}\otimes \mathcal{I}_{B}) \sqrt{\rho_{AB}}(\sigma_{jA}\otimes \mathcal{I}_{B})\big ],
          \end{equation}
 where i, j label the Pauli matrices.

Using above prescription, we can obtain the LQU for quantum state (\ref{reduced2}):
          \begin{equation}\label{LQU2}
\text{LQU}(\rho_{2}(t))=1-\text{max}\big(W_{1},W_{2}\big),
          \end{equation}
         where
                   \begin{equation}\label{W1}
         W_{1}=\sqrt {2e} \left( \sqrt {2\,e+r\sqrt {1-4\, \left( {\Gamma }^{
         2}-1 \right)  \left( p-1 \right) p}+r}+\sqrt {2\,e-r\sqrt {1-4\,
          \left( {\Gamma }^{2}-1 \right)  \left( p-1 \right) p}+r} \right),
                   \end{equation}
                   and 
                             \begin{equation}\label{W2}
                W_{2}=-{\frac {4\, \left( p-1 \right) p \left( r-2\,{\Gamma }^{2}\sqrt {{e}^
                {2}+er+ \left( {\Gamma }^{2}-1 \right)  \left( p-1 \right) p{r}^{2}}
                 \right) -8\, \left( {\Gamma }^{2}-2 \right) e \left( p-1 \right) p+4
                \,e+r}{4\, \left( {\Gamma }^{2}-1 \right)  \left( p-1 \right) p-1}}.
                             \end{equation}
\pagebreak

\end{document}